\documentclass[longauth]{aa}  

\usepackage{graphicx}
\usepackage{dblfloatfix}
\usepackage{caption}
\usepackage{subcaption}

\usepackage{xcolor}
\usepackage[normalem]{ulem}
\usepackage[switch]{lineno}
\usepackage{txfonts}
\usepackage{hyperref}
\hypersetup{
    colorlinks=true,
    linkcolor=blue,
    filecolor=magenta,      
    urlcolor=blue,
    citecolor=blue,
}
\usepackage{xspace} 
\newcommand{\nhitbin}{$f_{\rm Hit}$ bin\xspace}
\newcommand{\nhitbins}{$f_{\rm Hit}$ bins\xspace}
\newcommand{\nhit}{$f_{\rm Hit}$\xspace}
\newcommand{\gti}{GTI\xspace}
\newcommand{\gtis}{GTIs\xspace}
\newcommand{\healpix}{HEALPix\xspace}
\newcommand{\psf}{PSF\xspace}

\newcommand{\gammapy}{\textit{Gammapy}\xspace}
\newcommand{\hawc}{HAWC\xspace}
\newcommand{\iact}{IACT\xspace}
\newcommand{\iacts}{IACTs\xspace}
\newcommand{\cta}{CTA\xspace}
\newcommand{\swgo}{SWGO\xspace}
\newcommand{\lhaaso}{LHAASO\xspace}
\newcommand{\irf}{IRF\xspace}
\newcommand{\irfs}{IRFs\xspace}
\newcommand{\fermi}{\textit{Fermi}-LAT\xspace}
\newcommand{\gammaray}{$\gamma$-ray\xspace}

\newcommand{\gammahadron}{$\gamma$-hadron\xspace}

\newcommand{\gadf}{GADF\xspace}
\newcommand{\hess}{H.E.S.S.\xspace}

\begin{document} 

   \title{Validation of standardized data formats and tools for ground-level particle-based gamma-ray observatories.}

 \author{\small A.~Albert \inst{\ref{LANL}} \and R.~Alfaro \inst{\ref{IF-UNAM}} \and J.C.~Arteaga-Velázquez \inst{\ref{UMSNH}} \and H.A.~Ayala Solares \inst{\ref{PSU}} \and R.~Babu \inst{\ref{MTU}} \and E.~Belmont-Moreno \inst{\ref{IF-UNAM}} \and C.~Brisbois \inst{\ref{UMD}} \and K.S.~Caballero-Mora \inst{\ref{UNACH}} \and T.~Capistrán \inst{\ref{IA-UNAM}} \and A.~Carramiñana \inst{\ref{INAOE}} \and S.~Casanova \inst{\ref{IFJ-PAN}} \and O.~Chaparro-Amaro \inst{\ref{CIC-IPN}} \and U.~Cotti \inst{\ref{UMSNH}} \and J.~Cotzomi \inst{\ref{FCFM-BUAP}} \and S.~Coutiño de León \inst{\ref{UW-Madison}} \and E.~De la Fuente \inst{\ref{UdG}} \and R.~Diaz Hernandez \inst{\ref{INAOE}} \and M.A.~DuVernois \inst{\ref{UW-Madison}} \and M.~Durocher \inst{\ref{LANL}} \and C.~Espinoza \inst{\ref{IF-UNAM}} \and K.L.~Fan \inst{\ref{UMD}} \and M.~Fernández Alonso \inst{\ref{PSU}} \and N.~Fraija \inst{\ref{IA-UNAM}} \and J.A.~García-González \inst{\ref{ITESM}} \and H.~Goksu \inst{\ref{MPIK}} \and M.M.~González \inst{\ref{IA-UNAM}} \and J.A.~Goodman \inst{\ref{UMD}} \and J.P.~Harding \inst{\ref{LANL}} \and J.~Hinton \inst{\ref{MPIK}} \and D.~Huang \inst{\ref{MTU}} \and F.~Hueyotl-Zahuantitla \inst{\ref{UNACH}} \and P.~Hüntemeyer \inst{\ref{MTU}} \and A.~Jardin-Blicq \inst{\ref{armelle1},\ref{armelle2}, \ref{MPIK}} \and V.~Joshi \inst{\ref{ECAP}}{\large *} \and J.T.~Linnemann \inst{\ref{MSU}} \and A.L.~Longinotti \inst{\ref{IA-UNAM}} \and G.~Luis-Raya \inst{\ref{UPP}} \and K.~Malone \inst{\ref{LANL-Kelly}} \and V.~Marandon \inst{\ref{MPIK}} \and O.~Martinez \inst{\ref{FCFM-BUAP}} \and J.~Martínez-Castro \inst{\ref{CIC-IPN}} \and J.A.~Matthews \inst{\ref{UNM}} \and P.~Miranda-Romagnoli \inst{\ref{UAEH}} \and J.A.~Morales-Soto \inst{\ref{UMSNH}} \and E.~Moreno \inst{\ref{FCFM-BUAP}} \and M.~Mostafá \inst{\ref{PSU}} \and A.~Nayerhoda \inst{\ref{IFJ-PAN}} \and L.~Nellen \inst{\ref{ICN-UNAM}} \and M.U.~Nisa \inst{\ref{MSU}} \and R.~Noriega-Papaqui \inst{\ref{UAEH}} \and L.~Olivera-Nieto \inst{\ref{MPIK}}{\large *} \and E.G.~Pérez-Pérez \inst{\ref{UPP}} \and C.D.~Rho \inst{\ref{UOS}} \and D.~Rosa-González \inst{\ref{INAOE}} \and E.~Ruiz-Velasco \inst{\ref{MPIK}} \and D.~Salazar-Gallegos \inst{\ref{MSU}} \and F.~Salesa Greus \inst{\ref{IFJ-PAN}, \ref{CSIC}} \and A.~Sandoval \inst{\ref{IF-UNAM}} \and H.~Schoorlemmer \inst{\ref{MPIK}, \ref{Rab}}{\large *}  \and J.~Serna-Franco \inst{\ref{IF-UNAM}} \and A.J.~Smith \inst{\ref{UMD}} \and Y.~Son \inst{\ref{UOS}} \and R.W.~Springer \inst{\ref{University of Utah}} \and K.~Tollefson \inst{\ref{MSU}} \and I.~Torres \inst{\ref{INAOE}} \and R.~Torres-Escobedo \inst{\ref{SJTU}} \and R.~Turner \inst{\ref{MTU}} \and F.~Ureña-Mena \inst{\ref{INAOE}} \and L.~Villaseñor \inst{\ref{FCFM-BUAP}} \and X.~Wang \inst{\ref{MTU}} \and I.J.~Watson \inst{\ref{UOS}} \and E.~Willox \inst{\ref{UMD}} \and H.~Zhou \inst{\ref{SJTU}} \and C.~de León \inst{\ref{UMSNH}} \and  A.~Zepeda \inst{\ref{CINVESTAV}} (HAWC Collaboration) \and A.~Donath \inst{\ref{HARV}}{\large *} \and S.~Funk \inst{\ref{ECAP}}}

\institute{Physics Division, Los Alamos National Laboratory, Los Alamos, NM, USA  \label{LANL} \and
            Instituto de F\'isica, Universidad Nacional Autónoma de México, Ciudad de Mexico, Mexico  \label{IF-UNAM} \and 
            Universidad Michoacana de San Nicolás de Hidalgo, Morelia, Mexico  \label{UMSNH} \and 
            Department of Physics, Pennsylvania State University, University Park, PA, USA  \label{PSU} \and
            Department of Physics, Michigan Technological University, Houghton, MI, USA  \label{MTU} \and
            Department of Physics, University of Maryland, College Park, MD, USA  \label{UMD} \and 
            Universidad Autónoma de Chiapas, Tuxtla Gutiérrez, Chiapas, México \label{UNACH} \and 
            Instituto de Astronom\'ia, Universidad Nacional Autónoma de México, Ciudad de Mexico, Mexico  \label{IA-UNAM} \and 
            Instituto Nacional de Astrof\'isica, Óptica y Electrónica, Puebla, Mexico  \label{INAOE} \and 
            Institute of Nuclear Physics Polish Academy of Sciences, PL-31342 IFJ-PAN, Krakow, Poland  \label{IFJ-PAN} \and
            Centro de Investigaci\'on en Computaci\'on, Instituto Polit\'ecnico Nacional, M\'exico City, M\'exico. \label{CIC-IPN} \and 
            Facultad de Ciencias F\'isico Matemáticas, Benemérita Universidad Autónoma de Puebla, Puebla, Mexico  \label{FCFM-BUAP} \and 
            Department of Physics, University of Wisconsin-Madison, Madison, WI, USA  \label{UW-Madison} \and 
            Departamento de F\'isica, Centro Universitario de Ciencias Exactase Ingenierias, Universidad de Guadalajara, Guadalajara, Mexico  \label{UdG} \and Tecnol\'ogico de Monterrey, Escuela de Ingenier\'ia y Ciencias, Ave. Eugenio Garza Sada 2501, Monterrey, N.L., Mexico, 64849 \label{ITESM} \and 
            Max-Planck Institute for Nuclear Physics, 69117 Heidelberg, Germany \label{MPIK} \and 
            Department of Physics, Faculty of Science, Chulalongkorn University, 254 Phayathai Road, Pathumwan, Bangkok 10330, Thailand \label{armelle1}\and 
            National Astronomical Research Institute of Thailand (Public Organization), Don Kaeo, MaeRim, Chiang Mai 50180, Thailand \label{armelle2} \and 
            Erlangen Centre for Astroparticle Physics, Friedrich-Alexander-Universit\"at Erlangen-N\"urnberg, Erlangen, Germany \label{ECAP} \and
            Department of Physics and Astronomy, Michigan State University, East Lansing, MI, USA  \label{MSU} \and 
            Universidad Politecnica de Pachuca, Pachuca, Hgo, Mexico  \label{UPP} \and
            Space Science and Applications Group, Los Alamos National Laboratory, Los Alamos, NM, USA \label{LANL-Kelly} \and
            Dept of Physics and Astronomy, University of New Mexico, Albuquerque, NM, USA  \label{UNM} \and
            Universidad Autónoma del Estado de Hidalgo, Pachuca, Mexico \label{UAEH} \and
            Instituto de Ciencias Nucleares, Universidad Nacional Autónoma de Mexico, Ciudad de Mexico, Mexico  \label{ICN-UNAM} \and
            University of Seoul, Seoul, Rep. of Korea \label{UOS} \and 
            Instituto de Física Corpuscular, CSIC, Universitat de València, E-46980, Paterna, Valencia, Spain \label{CSIC} \and
            Radboud Universiteit, Nijmegen, Netherlands \label{Rab} \and
            Department of Physics and Astronomy, University of Utah, Salt Lake City, UT, USA  \label{University of Utah} \and
            Tsung-Dao Lee Institute and School of Physics and Astronomy, Shanghai Jiao Tong University, Shanghai, China \label{SJTU} \and
            Physics Department, Centro de Investigacion y de Estudios Avanzados del IPN, Mexico City, DF, Mexico \label{CINVESTAV} \and    
            Center for Astrophysics, Harvard and Smithsonian, Cambridge, MA, USA \label{HARV}
            }

\offprints{Laura Olivera-Nieto,
\protect\\\email{\href{mailto:laura.olivera-nieto@mpi-hd.mpg.de}{laura.olivera-nieto@mpi-hd.mpg.de}};
\protect\\\protect * Corresponding authors
}

   \date{Received March 11, 2022; accepted X XXX, 2022}
    \titlerunning{Validation of standardized data formats and tools for particle detector arrays.}
    \authorrunning{HAWC Collaboration}
  \abstract
   {Ground-based \gammaray astronomy is still a rather young field of research, with strong historical connections to particle physics. This is why most observations are conducted by experiments with proprietary data and analysis software, as is usual in the particle physics field. However, in recent years, this paradigm has been slowly shifting toward the development and use of open-source data formats and tools, driven by upcoming observatories such as the Cherenkov Telescope Array (CTA). In this context, a community-driven, shared data format (the \textit{gamma-astro-data-format,} or \gadf) and analysis tools such as \gammapy and \textit{ctools} have been developed. So far, these efforts have been led by the Imaging Atmospheric Cherenkov Telescope community, leaving out other types of ground-based \gammaray instruments.   }
   {We aim to show that the data from ground particle arrays, such as the High-Altitude Water Cherenkov (HAWC) observatory, are also compatible with the \gadf and can thus be fully analyzed using the related tools, in this case, \gammapy. }
   {We reproduced several published HAWC results using \gammapy and data products compliant with \gadf standard. We also illustrate the capabilities of the shared format and tools by producing a joint fit of the Crab spectrum including data from six different \gammaray experiments. }
   {We find excellent agreement with the reference results, a powerful confirmation of both the published results and the tools involved. }
   {The data from particle detector arrays such as the HAWC observatory can be adapted to the \gadf and thus analyzed with \gammapy. A common data format and shared analysis tools allow multi-instrument joint analysis and effective data sharing. To emphasize this, a sample of Crab nebula event lists is made public with this paper. 
   Because of the complementary nature of pointing and wide-field instruments, this synergy will be distinctly beneficial for the joint scientific exploitation of future observatories such as the Southern Wide-field Gamma-ray Observatory and CTA.
   
   }

   \keywords{Astronomical instrumentation, methods and techniques --
                methods: data analysis --
                Gamma rays: general
               }

   \maketitle
%
\section{Introduction}
In preparation for the upcoming Cherenkov Telescope Array (CTA), the ground-based \gammaray astronomy community has made a joint effort to define standardized data formats and develop community-sourced tools aimed to facilitate access to the data by a wide audience. This requires the identification of a data-processing stage in which standardization between different instruments is possible.
The primary source of background for any \gammaray observatory are events of hadronic origin usually referred to as cosmic rays. After reconstructing the events that triggered the detector, a background rejection step is applied in which \gammaray-like events are selected. At this stage, denoted as Data Level 3 (DL3) in the CTA data model \citep{data-levels}, the structure of the data of all \gammaray observatories is essentially the same. The DL3 is thus defined to include the $\gamma$-like event lists and the corresponding instrument response functions (\irfs). This  development and definition of a standard format for \gammaray astronomy has been a largely community-driven effort that is usually referred to as the \textit{gamma-astro-data-format,} or \gadf for short \citep{gadf-icrc17}. This format relies on file storage by the Flexible Image Transport System (FITS) format \citep{fits}, which is widely used by the whole astronomical community. It builds on existing standards such as the one developed by the FITS Working Group in the Office of Guest Investigators Program (OGIP) at NASA\footnote{\url{https://heasarc.gsfc.nasa.gov/docs/heasarc/ofwg/ofwg\_intro.html}} and expands them to address the specific needs of the \gammaray community. The availability of such an open data format will not only help to prepare the operation of CTA as an open observatory, but also simplify the process for existing observatories and experiments to possibly publish and archive their data in an openly documented and maintained data format.

With similar motivation, a variety of open-source analysis tools has been developed. This signals a transition in a field that up until recently, and with the notable exception of the \textit{Fermi} Large Area Telescope (LAT;~\citealt{fermipy}) or the \textit{INTEGRAL} analysis tools\footnote{\url{https://www.isdc.unige.ch/integral/analysis}}, for instance,
relied heavily on independent proprietary software developed for a specific observatory. These new open-source tools can be broadly classified into two classes. Some packages, such as the Multi-Mission Maximum Likelihood (3ML) \citep{3ml}, aim to bridge the gaps between different instruments by providing a common framework in which their respective proprietary tools interface, allowing joint, multiwavelength studies to be carried out. On the other hand, packages such as \gammapy \citep{gammapy-icrc17} and \textit{ctools} \citep{ctools} aim to replace the existing frameworks altogether, and offer a single tool with which to carry out the analysis of data from multiple \gammaray observatories, individually or jointly. 
The latter requires \gadf-conforming inputs, so that data from different observatories can be correctly read and analyzed by the same software.

There has been a number of studies that validated and highlighted the potential of the shared format and tools, focusing on either a single instrument \citep{lars} or on multi-instrument analysis \citep{cosimo}. 
However, these efforts have largely been focused on Imaging Atmospheric Cherenkov Telescopes (IACTs), excluding the other type of ground-based \gammaray instrument: particle detector arrays such as the High-Altitude Water Cherenkov (HAWC) observatory. While initially the focus of the \gadf and shared tools was on IACTs, given that CTA will be an array of such telescopes, the standard is in practice mostly compatible with the data from any other type of \gammaray instrument.

Because of the complementary nature of IACTs and particle detector arrays, including both in the conception and development of such tools can be very beneficial. Particle detector arrays continuously survey large fractions of the sky, but can do so with relatively low angular resolution~\citep{hawc-crab}. IACTs, on the other hand, have to be pointed to the region of interest, and are limited by weather and dark time, but can achieve higher angular resolution. \iacts can achieve good performance at low energies, below 1~TeV, while particle detector arrays are able to reach higher energies, of above 100~TeV. Multi-instrument analysis thus becomes necessary to cover the entire TeV range. A common data format and analysis tools allow the combination of data from \iact and particle detector arrays without the need for proprietary analysis software. This is relevant for both the current wide-field particle detector arrays, such as \hawc and the Large High Altitude Air Shower Observatory (\lhaaso;~\citealt{lhaaso}), and for future arrays such as the Southern Wide-Field Gamma-ray Observatory (SWGO;~\citealt{swgo}).

In this paper, we present the first full production of \hawc event lists and \irfs that follows the \gadf specification. We analyze it using \gammapy to reproduce a selection of published \hawc results. To do this, we start by building a background model that takes the produced event lists as input. Thereafter, we check the consistency of low-level data products such as the number of events and maps by comparing them with published examples. Furthermore, we reproduce three published \hawc results, each for a different source class by using~\gammapy. Last, as a proof of concept, we perform a joint fit using data of the Crab nebula from six different \gammaray observatories using~\gammapy.


\section{\hawc observatory}
\label{sec:hawc}
The High Altitude Water Cherenkov (\hawc) $\gamma$-ray observatory is situated on the flanks of the Sierra Negra at 18\degr59\arcmin41\arcsec{}N, 97\degr18\arcmin30.6\arcsec{}W in Mexico. It detects cosmic rays and $\gamma$-rays in the energy range from a few hundreds of GeV to more than a hundred TeV with a wide field of view (FoV) of $\sim2$ sr. \hawc has been fully operational with 300 Water Cherenkov Detectors (WCDs) since March 2015. In each such WCD of 4.5 m height and 7.3 m diameter, four photomultiplier tubes (PMTs) are submerged in 200 m$^3$ of purified water. The modular structure of \hawc WCDs allows optically isolating the detected Cherenkov light (300 to 500 nm) signal produced by the secondary particles such as e$^{\pm}$, $\gamma$, and $\mu^{\pm}$, while traveling through the water volume. It also facilitates the identification of the local variations in the observed lateral distribution of detected showers, which in turn greatly helps  performing \gammahadron separation.

The standard \hawc analysis procedure begins with the production of the instrument response functions (\irfs), which characterize the performance of the instrument. For this, air shower and detector simulations are generated using CORSIKA \citep{ref_CORSIKA} and a dedicated package based on GEANT4 (v4.10.00,~\citealt{ref_GEANT4}) named HAWCSim, respectively. These simulations are ran through the reconstruction procedure to obtain the two histograms that describe the detector response: the angular resolution and energy dispersion, the latter not normalized so that it also contains the effective area information. These quantities are usually stored in a ROOT~\citep{root} file.
\begin{table}
        \begin{center}
                \begin{tabular}{c c c}
                        \hline
                        \hline
                        Bin & Low edge (TeV) & High edge (TeV)\\
                        \hline
                        \hline
                        a &   0.316 &   0.562 \\
                        b &   0.562 &   1.00  \\
                        c &   1.00  &   1.78  \\
                        d &   1.78  &   3.16  \\
                        e &   3.16  &   5.62  \\
                        f &   5.62  &  10.0   \\
                        g &  10.0   &  17.8   \\
                        h &  17.8   &  31.6   \\
                        i &  31.6   &  56.2   \\
                        j &  56.2   & 100     \\
                        k & 100     & 177     \\
                        l & 177     & 316     \\
                        \hline
        \end{tabular} \end{center}
        \caption{Definition of the reconstructed energy bins. Each bin spans one quarter decade. The first two bins (a and b) are not used in the analysis as the estimate is highly biased.}
        \label{tab:bins}
\end{table}
\begin{table*}
\begin{center}
\begin{tabular}{ c c c c c c c c }
 \hline  \hline
Event ID & R.A & Dec. & Energy & Time & Core X & Core Y & Bin ID \\ 

 & \small (deg) & \small (deg) & \small (TeV) & \small (s) & \small (m) & \small (m) & \\ 
 \hline  \hline
1 & 296.401 & 18.649 & 6.698 & 1132183230.200404 & 50.4 & 212.8 & 7f \\
2 & 305.046 & 27.225 & 7.063 & 1132183236.213954 & -30.7 & 214.9 & 7f  \\
3 & 16.556 & 14.990 &  7.709 & 1132183250.7916136  &  -37.1 &  214.9 & 6f  \\
 \hline
\end{tabular}
\end{center}
\caption{Simplified entries of an event list. The real precision of the numbers has been reduced for formatting convenience. The Bin ID information is stored as an integer (e.g., bin 7f would correspond to number 77), with the bin name shown here only for convenience.}
\label{tab:events_example}
\end{table*}
 The reconstructed data are first binned depending on the fraction of the available PMTs triggered by the air shower, a quantity referred to here as \nhit. This results in a total of nine \nhitbins, referred to with integer numbers between 1 and 9, as described in~\cite{ref_hawc_crab_paper_2017}. The value of \nhit~is only weakly correlated with energy. In order to estimate the energy on an event-by-event basis, more advanced algorithms have been developed. The ground-parameter (GP) algorithm is based on the charge density deposited at the ground by the shower. The neural network (NN) algorithm estimates energies with an artificial neural network that takes as input several quantities computed during the event reconstruction. A detailed overview of both algorithms can be found in~\cite{hawc-crab}. All results shown in this paper correspond to energies estimated using the GP method, but that is only for simplicity, as it is also possible to use the NN estimator results instead. 

Energy bins are usually defined beforehand, with 12 reconstructed energy bins, each spanning a quarter decade in $\log_{10}(E/\rm{TeV})$. Energy bins are labeled alphabetically with increasing energy, as shown in Table~\ref{tab:bins}. The combination of both binning schemes leads to a total of 108 2D \nhit/energy bins~\citep{hawc-crab}, identified by the combination of the \nhit number and the energy letter.
For each bin, the \gammahadron separation cuts are optimized independently and applied to the reconstructed data. A detailed description of the variables used for \gammahadron separation can be found in~\cite{ref_hawc_crab_paper_2017}. 

DL3 products are currently not produced during the \hawc standard analysis procedure, and instead, the events are selected for $\gamma$-likeness and directly binned into a \healpix~\citep{healpix_paper} full-sky map. In this step, a pointing correction, usually referred to as alignment, is applied to the reconstructed data as described in~\cite{ref_hawc_crab_paper_2017}. During the same map-making procedure, a background and exposure map is computed as well~\citep{ref_hawc_crab_paper_2017}. These maps and the detector response file are typically used within 3ML~\citep{3ml} via the \hawc-specific plugin \texttt{hawc\_hal}~\citep{hawchal} to carry out \gammaray source analysis.

\section{\gammapy}
\label{sec:gammapy}

\gammapy is a community-developed Python package for \gammaray astronomy. It is built on the scientific Python standard packages \textit{Numpy}, \textit{Scipy}, and \textit{Astropy} and implements data reduction and analysis methods for \gammaray astronomy. It will also be used as the base package for the science tools for the future \cta. \gammapy has already been successfully used and validated for analysis of \iact data from \hess~\citep{lars} and has also been used to perform joint analyses of multiple \iacts with \fermi ~\citep{cosimo}. The standard analysis workflow of \gammapy begins at the level of selecting the DL3 data and time intervals based on Good Time Intervals (\gtis). In the next step, selected events are binned into multidimensional sky maps, such as the World Coordinate System (WCS) or \healpix with an additional energy axis. The corresponding instrument response, including the residual hadronic background, is projected onto the same but possibly spatially coarser sky map. The binned data are bundled into a dataset, and together with a parametric model description, they can be used to model the data in a binned likelihood fit. Multiple datasets can be combined, and by sharing the same source model, they can be used to handle multiple event types or data from different instruments in a joint-likelihood fit. To fully support the analysis of \hawc data, we made one contribution to \gammapy. The \hawc point-spread function (\psf) is computed as a function of reconstructed energy, as opposed to true energy, which is typical for IACTs. For this reason, we implemented the possibility to exchange the order
of the application of the \psf and energy dispersion matrix (see Section~\ref{subsec:irfs}). All of the other features are already compatible with standard analysis workflows used for ground-based wide-field instruments. This includes combined spectral and morphological modeling of \gammaray sources, computation of test-statistic (TS) maps, and estimation of flux points and light curves. All the results shown in this paper were produced using \gammapy version \texttt{0.18.2}.

\section{\hawc data and \irfs in the \gadf}
\label{sec:dl3}
At the DL3 level, the GDAF defines mandatory header keywords and columns, containing the basic information necessary for \gammaray data analysis, as well as optional entries that can be useful for specific instruments or observing strategies. The storage and distribution of \gammaray data as event lists together with some parameterization of the \irfs has been shown to be extremely successful by the \fermi observatory\footnote{\url{https://fermi.gsfc.nasa.gov/cgi-bin/ssc/LAT/LATDataQuery.cgi}}. This format was extended to \iacts by the \gadf initiative, and is extended in this work to particle detector arrays such as the \hawc Observatory.

\subsection{Event lists}
\label{subsec:events}

The DL3 stage refers to lists of reconstructed events that have been identified as $\gamma$-ray-like. Right after reconstruction, \hawc events are stored in event lists that mostly contain background events of a hadronic nature. The first step toward DL3 event lists is thus to bin them as described in Section~\ref{sec:hawc} and apply the corresponding \gammahadron separation criteria in each bin to select \gammaray-like events. An additional alignment correction is applied to the direction of each event (see~\citealt{ref_hawc_crab_paper_2017}), and the coordinates are transformed into the J2000 epoch.
For each of the events, five basic quantities are required by the \gadf: an event identification number, the two sky direction coordinates, the estimated energy, and the arrival time~\citep{gadf-icrc17}. Event time-stamps are stored in GPS seconds after midnight January 6, 1980\footnote{\url{https://gssc.esa.int/navipedia/index.php/Time_References_in_GNSS}}, with the reference time provided in the FITS file header.  Additionally, any other instrument-specific column can be added, such as the fraction of available triggered PMTs or the core location of the shower in the detector coordinates. Table~\ref{tab:events_example} shows an example of such an event list. An integer indicating to which of the 108 2D bins each event belongs to is added as a column. This allows storing events from different bins in the same file without any loss of information.

\subsection{GTIs and exposure calculation}
\label{subsec:gtis}

\begin{figure}[]
        \includegraphics[width=0.46\textwidth]{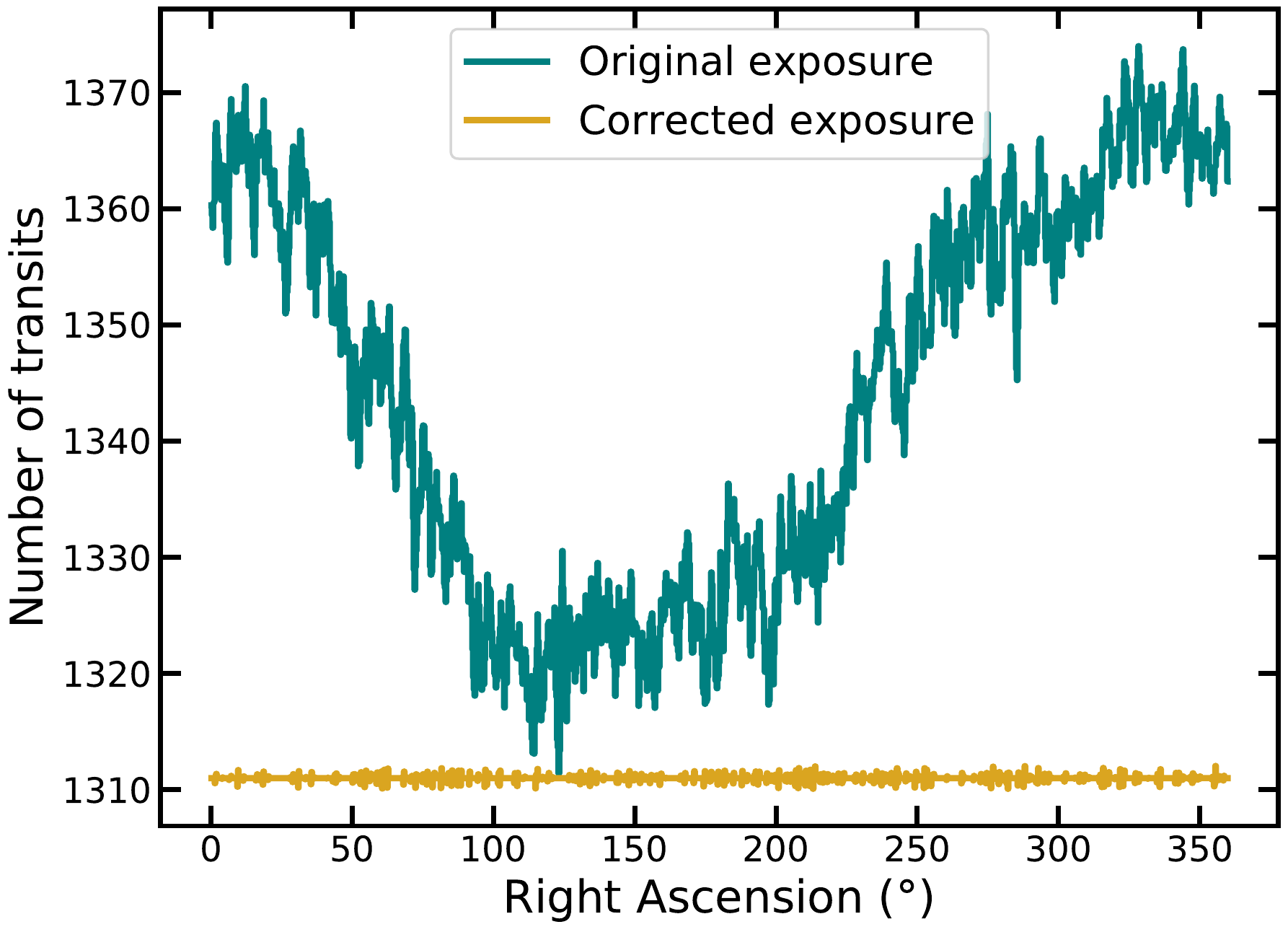}
        \caption{Number of transits during which the detector was stable as a function of R.A.}
        \label{fig:livetime}
\end{figure}

The \gtis are defined as the time intervals during which the detector is stable and taking data. They are stored as a separate table within the same FITS file as the associated event list. They are used to compute the exposure time
, which is crucial for measuring, for example, the \gammaray flux of astrophysical sources.

HAWC raw data are stored in files that span 125~s of data-taking. After the data are reconstructed, these intervals are checked for stability~\citep{hawc-crab}; their duration becomes the minimum unit of time that can be described as good. The \gadf requires the \gti table to have two columns, one with the start of the interval (TSTART), and one with the end (TSTOP). For the DL3 production presented here, these time stamps are obtained from the reconstructed data files themselves. This is done by selecting the first and last event in a file before applying any \gammahadron separation or binning. 
Currently, the effect of trigger dead time, which is expected to be in the order of a few percent, is not taken into account when analyzing HAWC data.
From these time intervals, the exposure map is constructed by considering which part of the sky is above the maximum zenith angle observation threshold as seen from the observatory during each interval.

Because of the continuous observations performed by \hawc, it is often useful to describe exposure in terms of a source transit above the detector, which is defined in~\cite{hawc-lightcurve}. The green curve in Figure~\ref{fig:livetime} shows the number of transits during which \hawc was stable and taking data between June 2015 and June 2019 as a function of right ascension (R.A.). Detector downtime can be caused by a variety of factors, ranging from hardware issues to meteorological conditions such as electric storms. As a result, these interruptions are not necessarily distributed uniformly over time; technical maintenance, for instance, is more likely during particular times of the day. This leads to the fluctuations in the exposure as a function of hour angle, or equivalently, R.A., that is shown as the green curve in Figure~\ref{fig:livetime}. These fluctuations are on the 3\% level and are usually neglected in long-term source analysis. One of the advantages of the production of \gtis together with event lists is that this effect becomes easy to characterize and correct for. 

The different data ranges defined by the \gtis can still be ranked by detector stability criteria, and those ranked lower are iteratively removed, effectively shaving time off of the green curve in Figure~\ref{fig:livetime} until it becomes flat. The result is shown by the curve labeled "Corrected exposure" in Figure~\ref{fig:livetime} . This allows accurately neglecting the R.A dependence of the live time while still keeping a total data efficiency of more than 90\%.

\subsection{Instrument response}
\label{subsec:irfs}
The \irfs describe the combined detection abilities and precision of an instrument data-taking and reconstruction procedure.
Independent of the actual detection principle, the response of a \gammaray instrument can be described by a few key properties. The angular resolution of the experiment is the reconstruction accuracy of the direction of the incident \gammaray, and is described by the point-spread function ($PSF$). The energy dispersion ($E_{\rm disp}$) is the reconstruction accuracy of the energy of each event. The detection probability of a \gammaray is given by the effective area ($A_{\rm eff}$). Finally, the expected residual hadronic background by misclassified events ($N_{\rm B}$) is described by the background model (see Section \ref{sec:background}).

The current version of the \gadf neglects the correlation between $PSF$, $E_{\rm disp}$, and $A_{\rm eff}$ and considers them independent. This can be described by the following combined instrument response~$R$:
\begin{equation}
R(\mathbf{x}, E|\mathbf{x'}, E') = A_{\rm eff}(\mathbf{x'}, E') \cdot PSF(\mathbf{x}|\mathbf{x'}, E) \cdot E_{\rm disp}(E|\mathbf{x'}, E')
\label{eq:irf_factorization}
,\end{equation}

where $\mathbf{x}$ and $E$ represent the reconstructed position and energy, while $\mathbf{x'}$ and $\mathbf{E'}$ are the corresponding unknown true quantities. The assumption of independence  is mostly sufficient for the current generation of instruments, including \hawc. However, it will be readdressed for \cta and likely \gadf in the future. As mentioned in Section~\ref{sec:gammapy}, the \hawc PSF is currently provided in reconstructed energy, which introduces a dependence on the assumed spectral index of the modeled source. However, the data format also allows defining PSF in true energy as well, which  allows the spectral reweighting of the PSF during model evaluation. Using this assumption, predicted counts N$_{\rm Pred}$ can be computed as
\begin{equation}
   {N}(\mathbf{x}, E) = {N}_{\rm B}(\mathbf{x}, E) + t_{\rm live} \int_{\mathbf{x'}} {\rm d}\mathbf{x'} \, \int_{E'} {\rm d}E' \,  R(\mathbf{x}, E|\mathbf{x'}, E') \cdot \Phi(\mathbf{x'}, E')
\label{eq:npred_model}
,\end{equation}
where N$_{\rm B}$ is the expected residual hadronic background, $t_{\rm life}$ is the live time, $R$ is the combined instrument response as defined in Equation~\ref{eq:irf_factorization} and $\Phi(\mathbf{p'}, E')$ the flux of the source model. 
One set of each \irf is produced per analysis bin because they are independent, resulting in a value of N$_{\rm Pred}$ per analysis bin. More details of the \hawc \irfs in the \gadf can be found in~\cite{the-proceeding}.

\section{Background modeling}
\label{sec:background}

\subsection{Derivation of the background model}
\label{subsec:background-constr}

\begin{figure*} []
        \centering
        \begin{subfigure}[b]{0.3\textwidth}
                \centering
                \includegraphics[width=\textwidth]{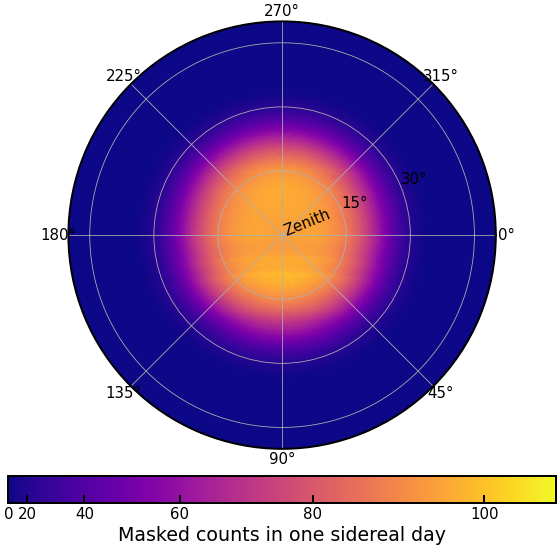}
        \end{subfigure}
        \begin{subfigure}[b]{0.3\textwidth}
                \centering
                \includegraphics[width=\textwidth]{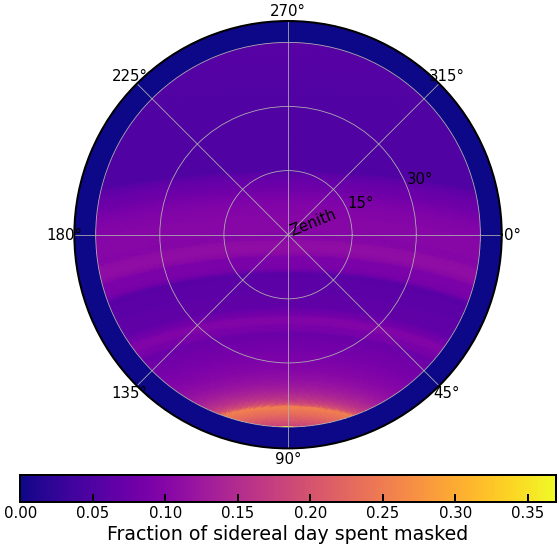}
        \end{subfigure}
        \begin{subfigure}[b]{0.3\textwidth}
                \centering
                \includegraphics[width=\textwidth]{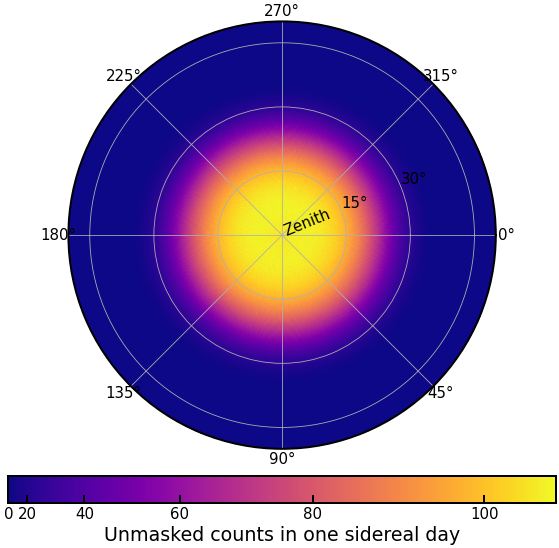}
        \end{subfigure}
        \caption{Local coordinates view of the different quantities relevant to the background model construction.     	
        	\textit{Left:} Masked spatial template for bin 1c, $\bar{B}_{M}(\theta, \phi)$. \textit{Middle:} Mask weights, $W_M(\theta, \phi)$, quantifying the fraction of the total time that a pixel is masked. \textit{Right:} Weighted (unmasked) spatial template for bin 1c, $\bar{B}(\theta, \phi)$.}
\label{fig:spatial_template}
\end{figure*}
Background estimation in \hawc analysis is typically done using the so-called direct integration method~\citep{milagro}. This method deals with the expected dipole cosmic-ray anisotropy by splitting the data into time intervals (usually 2h) and estimating the background in each of these intervals, which are then added up. This requires the input files to be provided chronologically sorted and is typically done in the same process as the \gammahadron separation and map-making. However, the production of $\gamma$-like event lists simplifies this process by allowing the use of the entire dataset at once with the slightly modified method described below. This has the advantage of significantly reducing the required computing time, given that the input events are already selected as \gammaray-like, as well as providing additional flexibility and modularity to the process. Removing the need for small sequential time intervals also leads to improved statistics at the highest energies.


At a given sidereal time, every day, the region of sky above the observatory is the same. The events in the event lists were selected using the \gtis described in Section~\ref{subsec:gtis}, and split into 720 bins of sidereal time, $\tau$. The duration of the bins is thus chosen to be 2~min of sidereal time, during which the sky above the observatory moves by 0.5º. This very fine binning is helpful to account for the dipole anisotropy. In each of these sidereal time bins, a sky map in local coordinates was filled using the selected events for each of the 2D analysis bins introduced in Section~\ref{sec:hawc}, which we refer to as $B_{\tau}(\theta, \phi),$ where $\theta$ and $\phi$ are the zenith and azimuth angles, respectively.
We define a mask to exclude a band of $\pm$4\degr around the Galactic plane, as well as other known bright \gammaray sources, as detailed in Table~\ref{tab:mask}. We computed the mask in local coordinates for each of the defined sidereal time intervals, $M_{\tau}(\theta, \phi)$.
\begin{table}[h!]
\begin{center}
\begin{tabular}{c c c c}
 \hline  \hline

Component & Center (lº, bº) & Shape & Width/Radius (º) \\ 
 \hline
 \hline

Galactic Plane & (0, 0) & Band & 8 \\ 
Geminga & (195.14, 4.27) & Disk & 10  \\  
Monogem & (201.11, 8.26) & Disk & 10  \\ 
Mrk421 & (179.88, 65.01) & Disk & 2  \\  
Mrk501 & (63.60, 38.86) & Disk & 2  \\  
Crab & (184.56, -5.78) & Disk & 2 \\  
\hline
\end{tabular}
\end{center}
\caption{Mask components. The center of the region is given in Galactic coordinates.}
 \label{tab:mask}
\end{table}

\begin{figure}[]
        \includegraphics[width=0.49\textwidth]{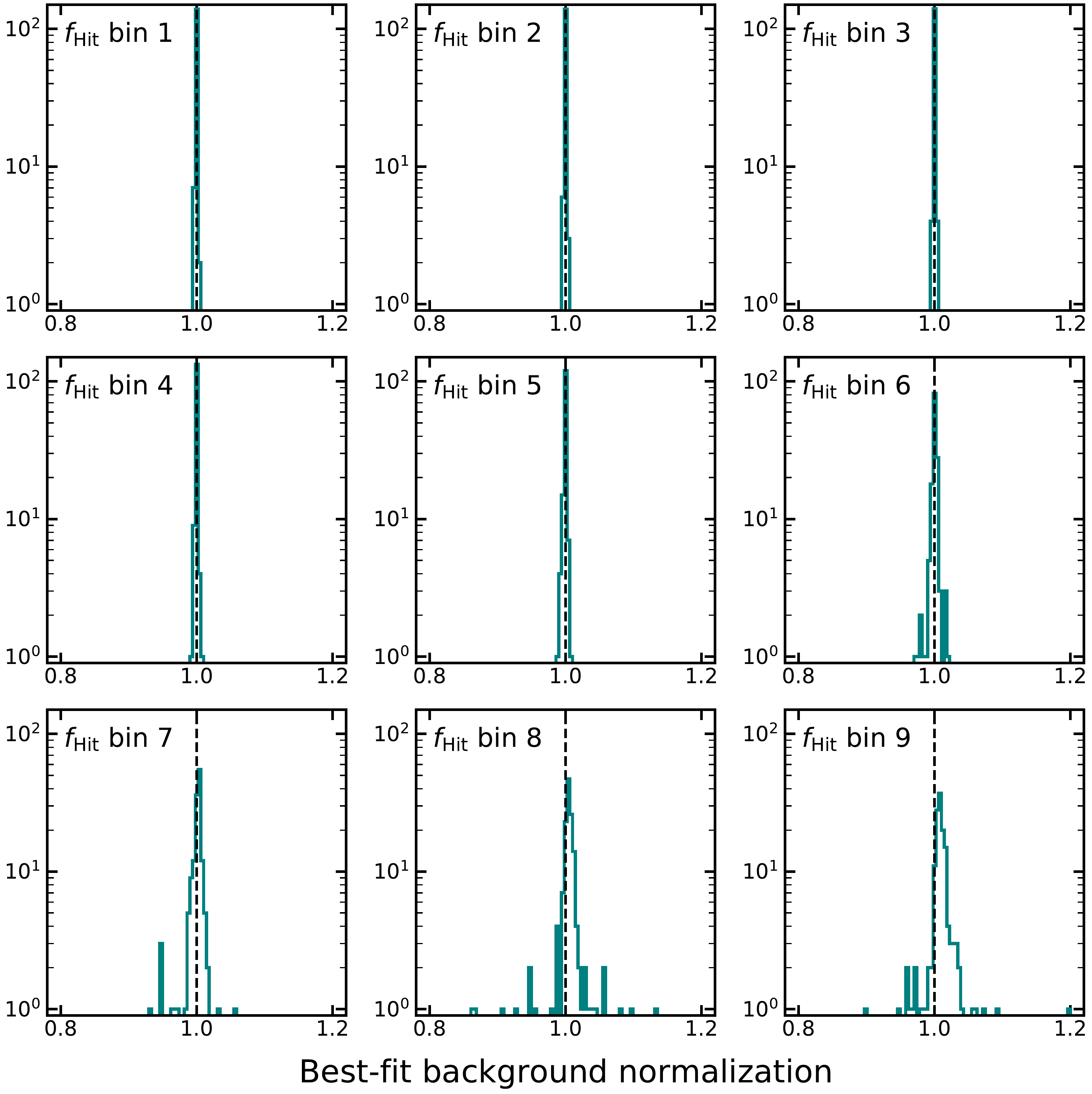}
        \caption{Best-fit results for the background normalization of the tiles for each of the nine \nhitbins. }
        \label{fig:bkgchecknorm}
\end{figure}
\begin{figure*}[b]
    \centering
                \includegraphics[width=0.75\textwidth]{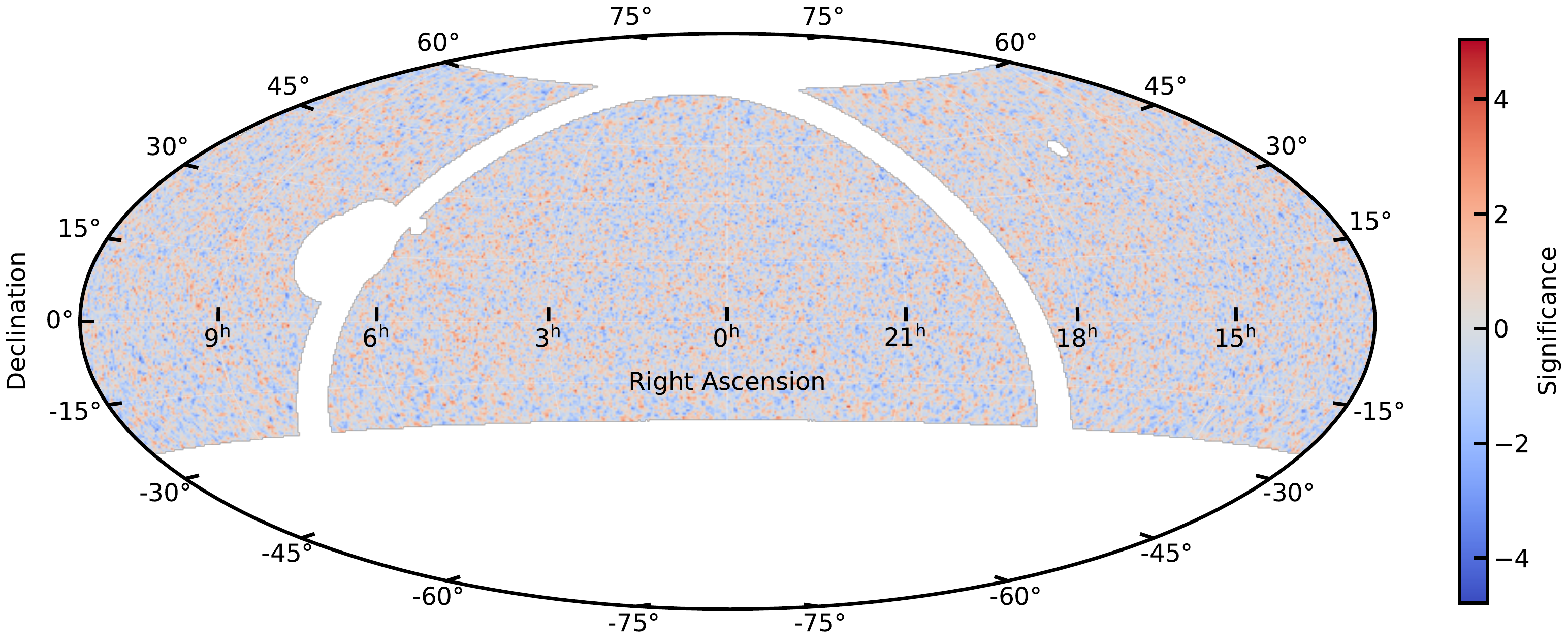}

        \caption{Full-sky significance map for \nhit bin 4 as computed with \gammapy. The map is masked using the mask described in Table~\ref{tab:mask}. }
        \label{fig:sig_map}
\end{figure*}

From these ingredients, we can construct the background model. First, we mask and add the maps in sidereal time to build a time-independent masked spatial template,
\begin{equation}
\bar{B}_{M}(\theta, \phi) = \sum_{\tau} M_{\tau}(\theta, \phi) \cdot B_{\tau}(\theta, \phi).
\end{equation}
In order to correct for the presence of the mask, we integrate the mask in sidereal time to compute weights quantifying the fraction of a sidereal day that each spatial pixel spends inside of the mask,
\begin{equation}
W_M(\theta, \phi) = \left( \sum_{\tau} M_{\tau}(\theta, \phi) d\tau \right)^{-1}.
\end{equation}

We can now recover the unmasked spatial templates,
\begin{equation}
\bar{B}(\theta, \phi) =W_M(\theta, \phi) \cdot \bar{B}_{M}(\theta, \phi).
\end{equation}
An example of this process is shown in Figure~\ref{fig:spatial_template}.

These spatial templates represent the time-independent spatial distribution of events in the \hawc sky for each of the 2D analysis bins. To account for temporal fluctuations in the event rate, we compute the event rate outside of the exclusion mask for the maps in sidereal time bins,
\begin{equation}
R(\tau) = \sum_{\theta, \phi} M_{\tau}(\theta, \phi) \cdot B_{\tau}(\theta, \phi)
,\end{equation}
and the event rate outside of the exclusion mask in the spatial template,
\begin{equation}
\bar{R}(\tau) = \sum_{\theta, \phi} M_{\tau}(\theta, \phi) \cdot \bar{B}(\theta, \phi).
\end{equation}
We can now combine the time-independent spatial template $\bar{B}(\theta, \phi)$ for each analysis bin with the time-dependent rate as
\begin{equation}
B(\theta, \phi, \tau) = \bar{B}(\theta, \phi) \cdot \frac{R(\tau)}{\bar{R}(\tau)}.
\end{equation}
This results in $B(\theta, \phi, \tau)$, the background map in local coordinates for each sidereal time interval, which takes into account the fluctuations in the event rate. Finally, in order to construct the desired map in sky coordinates, we project each of the local coordinate maps corresponding to a sidereal time $\tau$ into the corresponding sky coordinates and stack them together into one full-sky map. This process yields one such map per analysis bin. The different energy bins can be bundled together into groups of the same \nhitbin, which results in a three-dimensional sky map that includes an energy axis for each of the \nhitbins.

\subsection{Checks of the background model}

\subsubsection{Background normalization}
\begin{figure}
\centering
        \includegraphics[width=0.45\textwidth]{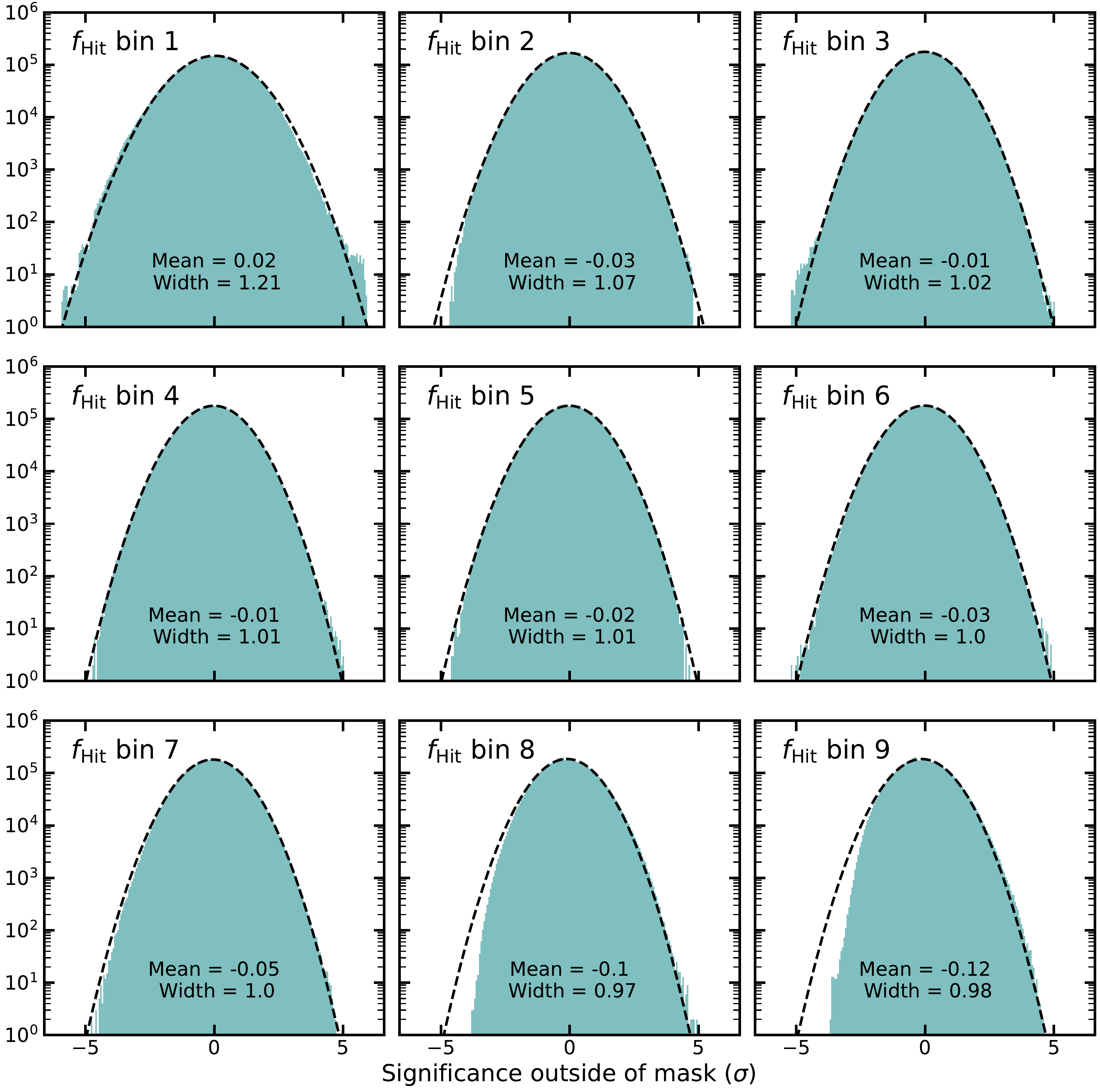}
        \caption{Distribution of the significance values outside of the mask for all the \nhit bins (green). A Gaussian function is fit to each of the histograms (dashed black line), and the best-fit mean and width are given in each panel.}
        \label{fig:significance}
\end{figure}
\begin{figure}
        \includegraphics[width=0.48\textwidth]{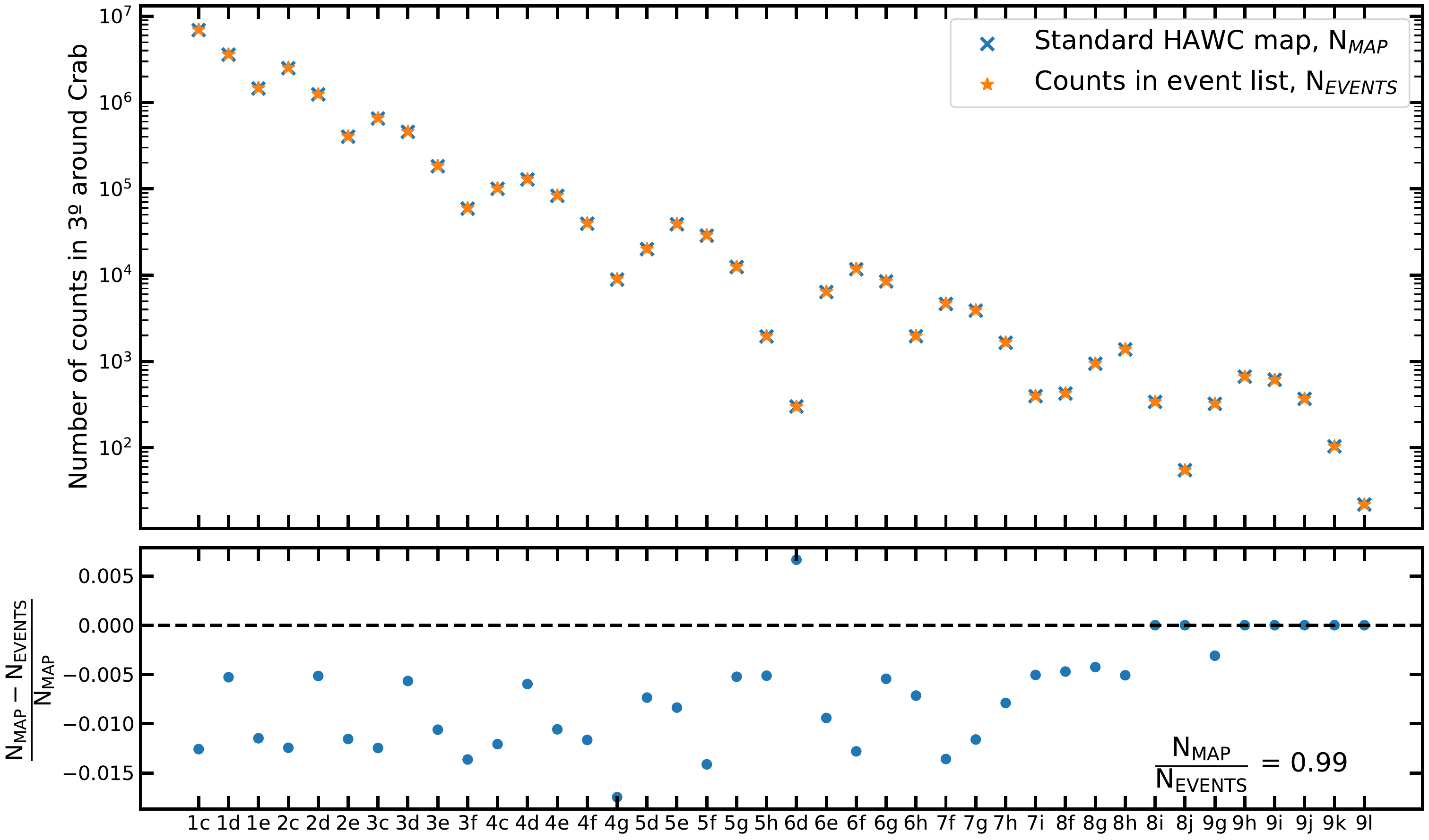}
        \caption{Comparison of the number of events in a region of 3\degr~radius around the Crab nebula in the standard \hawc map and in the event lists for each of the analysis bins. The selected 2D bins shown here are those that are used in Section~\ref{subsec:crab} and follow the selection procedure described in~\cite{hawc-crab}.}
        \label{fig:count_comparison}
\end{figure}

\begin{figure*}
        \centering
        \begin{subfigure}[b]{0.3\textwidth}
                \centering
                \includegraphics[width=\textwidth]{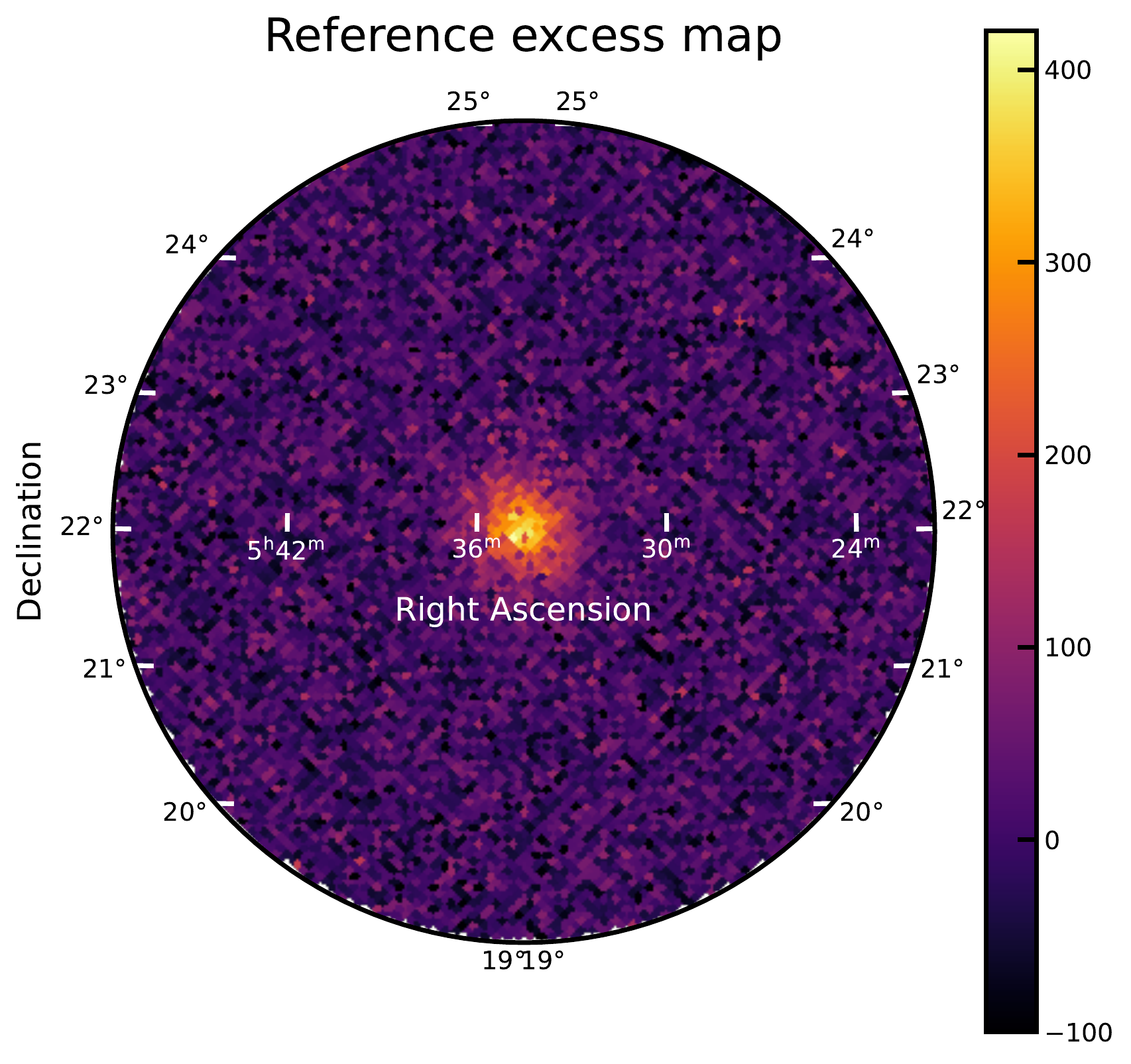}
        \end{subfigure}
        \begin{subfigure}[b]{0.3\textwidth}
                \centering
                \includegraphics[width=\textwidth]{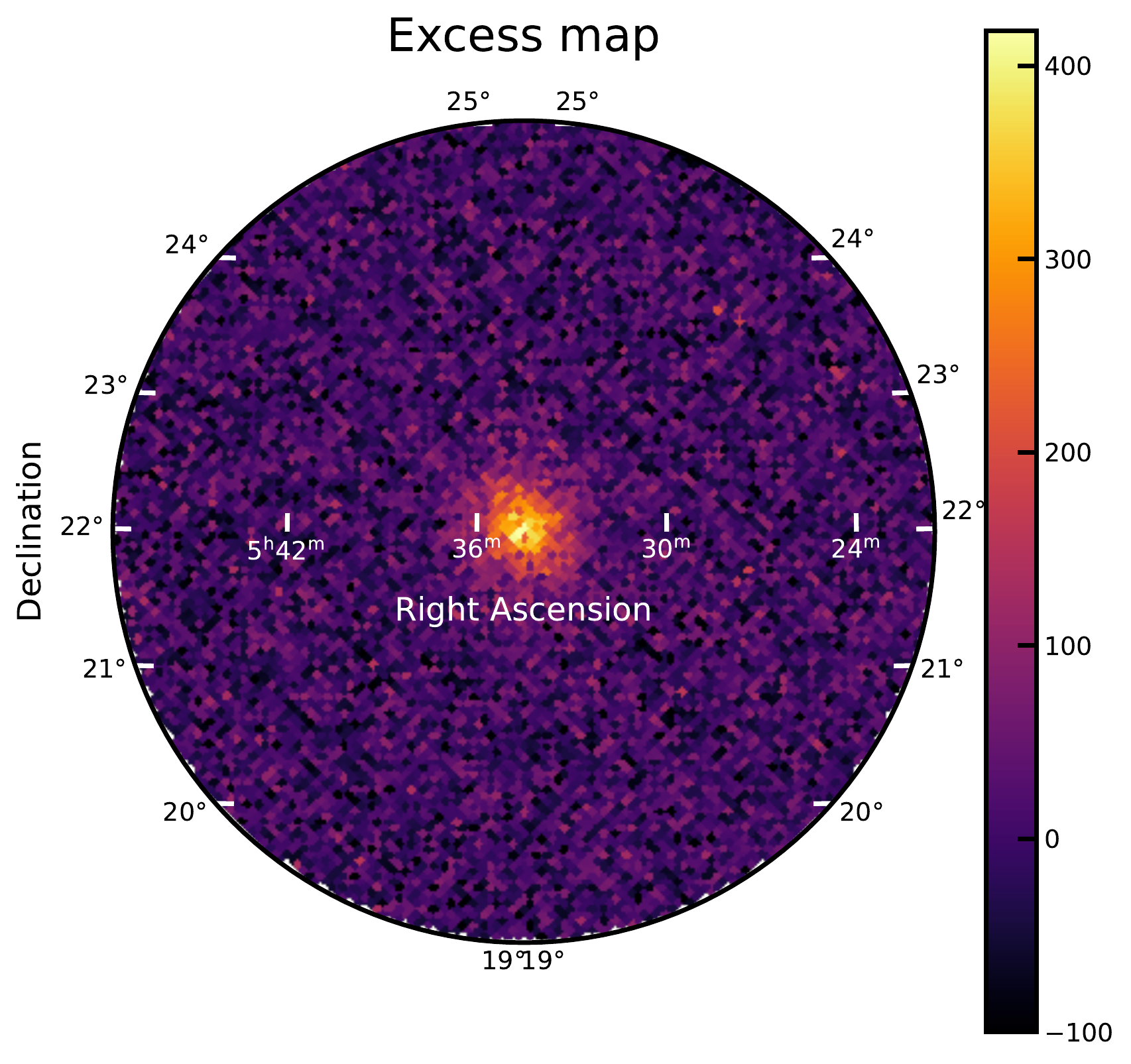}
        \end{subfigure}
        \begin{subfigure}[b]{0.3\textwidth}
                \centering
                \includegraphics[width=\textwidth]{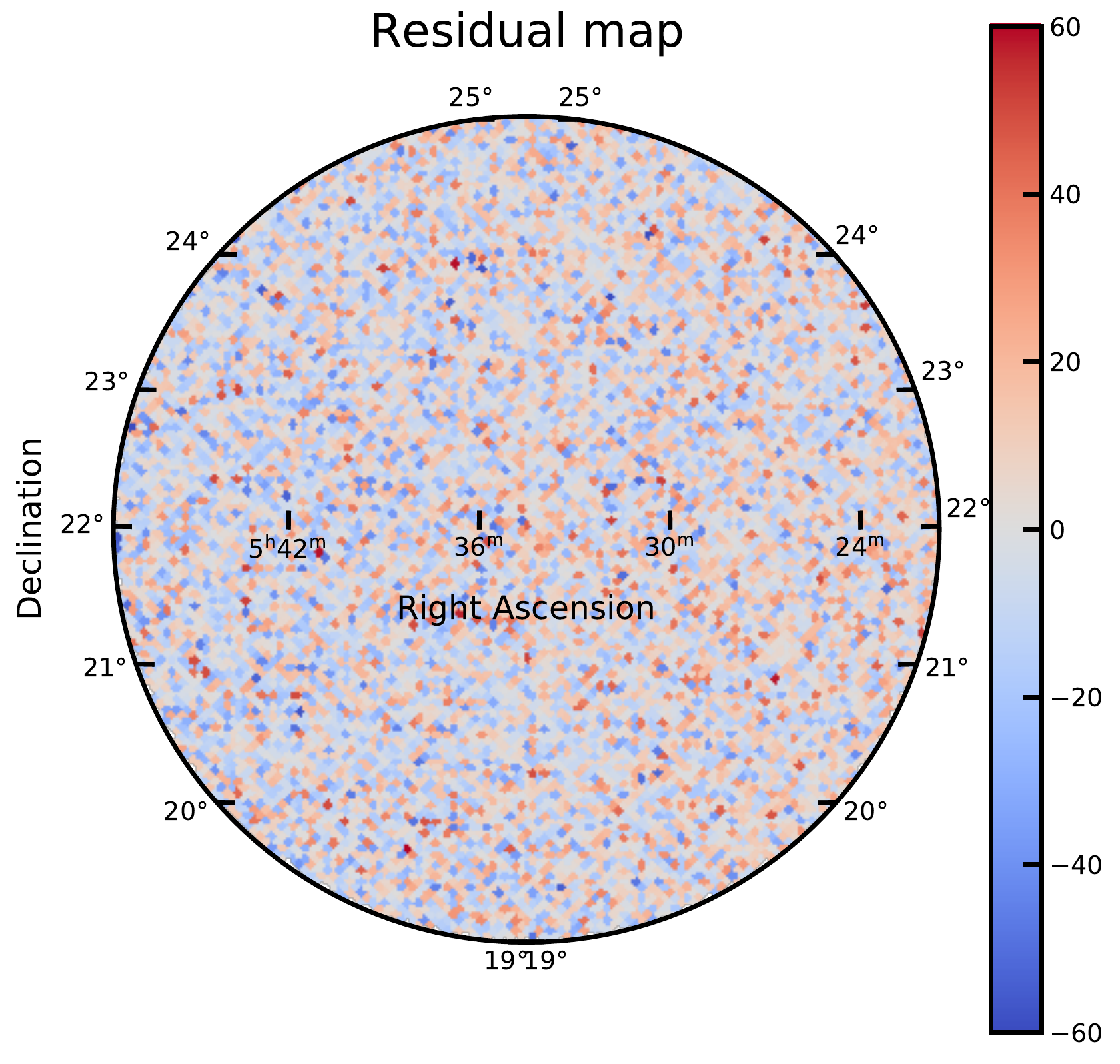}
        \end{subfigure}
        \caption{Comparison of excess counts maps of the Crab nebula region.       	
        	\textit{Left:} Crab excess counts map above 1 TeV as computed from the standard HAWC pipeline. \textit{Middle:} Crab excess counts map above 1 TeV as computed from the DL3 data products. \textit{Right:} Residual map resulting from subtracting the reference map from the map derived from the DL3 data products.}
\label{fig:excess_map}
\end{figure*}
To validate the background model, we split the full-sky counts map into 192 tiles of equal solid angle, applying the same mask as was used for the background model creation. For the 148 tiles that are at least partially contained in the \hawc FoV, we then compared the background model to the observed counts outside of the mask, where no bright \gammaray sources are expected. To do this, we defined a background normalization parameter that multiplies the background model, and performed a fit. Figure~\ref{fig:bkgchecknorm} shows the histogram of the resulting best-fit background normalization for the tiles. The normalization distributions are centered around 1, as is expected for a well-normalized background model.


\subsubsection{Full-sky significance map}

Because particle detector arrays continuously survey large fractions of the sky, producing full-sky maps is critical for the science and diagnosis of the data products. An example of this are full-sky significance maps, which can also be used to identify new sources in the instrument FoV. Such a map has been shown repeatedly by the \hawc Collaboration, for example, in~\cite{3HWC-catalog}. Using the background map produced as described in Section~\ref{subsec:background-constr} for 1311 transits and the corresponding count map produced with the event list, we can compute the significance map using \textit{Gammapy}. We used this map to confirm the background model because we expect the significance to have no hotspots above 5$\sigma$ and to be normally distributed outside of the mask described in Table~\ref{tab:mask}. The general approach to this is again to divide the HEALPix-based all-sky data into smaller patches using tangential WCS projections, compute the maps, and reproject back to a HEALPix pixelization. One of the resulting maps, for \nhit bin 4, is shown in Figure~\ref{fig:sig_map}. A histogram of the masked significance for all the other bins is shown in Figure~\ref{fig:significance}. As expected, there are no regions in the map with a significance above 5$\sigma$. The significance histograms for most bins follow a Gaussian distribution with a mean value of roughly zero and a width of unity, as expected from random fluctuations. The broader distribution in bin 1 is due to the imperfect characterization of the cosmic-ray anisotropy, which is most relevant in bins in which the background rate is higher, that is, low \nhitbins. Additionally, all the pixels with a significance above 5$\sigma$ in \nhitbins 1 and 3 are located close to the edge of the HAWC FoV, indicating that they are likely the result of an edge effect of the map. The deviation from Gaussian behavior in bins 8 and 9 is explained by the relatively low number of events in these bins. Any source that is not covered by the mask is expected to contribute to the distributions shown in Figure~\ref{fig:significance}. However, following the construction of the mask, these sources would be faint, meaning that their individual contribution to each \nhitbin is unlikely to cross the 5$\sigma$ threshold.

\section{Comparison of data products}
\label{sec:dl4_validation}

In order to ensure that the event selection and alignment were performed correctly, we can compare the number of events classified into each bin. Because event lists were not previously produced in \hawc, we do this by comparing the number of counts in the standard maps to the number of events in the lists for the region defined by a radius of 3º  around the Crab nebula. We make this comparison prior to the exposure flattening described in Section~\ref{subsec:gtis} in order to compare the same number of data. We expect the event lists to contain slightly more events than the maps because a few percent of the total events is rejected during the standard map-making process due to criteria on the gaps between the time-stamp of events required by the exposure calculation. This is no longer necessary when the exposure is computed by using \gtis, as described in Section~\ref{subsec:gtis}. This effect is larger for bins with more events, such as low \nhitbins.   Figure~\ref{fig:count_comparison} shows that the number of counts agrees well for all bins. The total difference is about the expected 1\%.

In addition to the total number of events, it is important to also ensure that their spatial distribution follows the expectations. Figure~1 in~\cite{hawc-crab} shows the excess map of the region around the Crab nebula above 1~TeV for 837.2 days. We reproduced this excess map for the same data range and present both maps together with the residual between them in Figure~\ref{fig:excess_map}. It is clear that the maps are strikingly similar.

\section{Validation}
\label{sec:validation}
To validate the production of the event lists and IRFs, we chose three sources representing three different analysis use cases: the Crab nebula (eHWC J0534+220) as the standard candle and Galactic point source, the extended source eHWC J1907+063, and the extragalactic variable source Mrk 421.

For the two steady source analyses, we used the framework described previously, with the events and \irfs described in Section~\ref{sec:dl3} and the background model described in Section~\ref{sec:background}. For the special case of Mrk 421, the background was estimated locally, as detailed in Section~\ref{subsec:mrk421}. Despite this difference, the workflow was very similar in all three cases. Events were selected from a 8\degr$\times$8\degr~region in the sky around the source position. For each of the \nhitbins described in Section~\ref{sec:hawc}, a three-dimensional map was produced, with two spatial axes and a reconstructed energy axis, binned as also described in Section~\ref{sec:hawc}. The background map was interpolated to that same geometry, and so were the \irfs described in Section~\ref{sec:dl3}. As an additional check and because it is also possible within \textit{Gammapy}, the analyses in Sections~\ref{subsec:crab} and~\ref{subsec:j1908} were also carried out using the existing \hawc counts and direct integration background maps.

The data and \irfs were bundled into a \gammapy dataset (see Section~\ref{sec:gammapy}), one for each \nhitbin. Then, the relevant model was attached to the datasets, and a joint-likelihood fit was performed to all nine datasets together. The only difference in the case of Mrk 421 is that this same procedure was carried out for each of the selected time intervals to build the light curve.

We do not expect to exactly reproduce the reference best-fit values for several reasons. First, some of the validation analyses shown here make use of a different background model than the reference (see Section~\ref{sec:background}). Second, as mentioned in Section~\ref{subsec:gtis}, the exposure for the reference analyses is assumed to be flat. This introduces an error in the flux of up to a few percent that is not present in the validation analysis. Finally, the data reduction process described in Section~\ref{sec:gammapy} includes the projection into a sky-map and interpolation of the \irfs to a coordinate grid centered on the source position. This is not done in the reference analysis, which uses the \irf value that corresponds to the nearest declination node, spaced by 5\degr. This can result in differences for the best-fit parameters, especially for sources located between declination nodes for which the \irfs evolve rapidly in the spatial dimension. All error values shown throughout this section are statistical only.

\subsection{Point source: Crab nebula}
\label{subsec:crab}
\begin{figure}[]
\centering
        \includegraphics[width=0.46\textwidth]{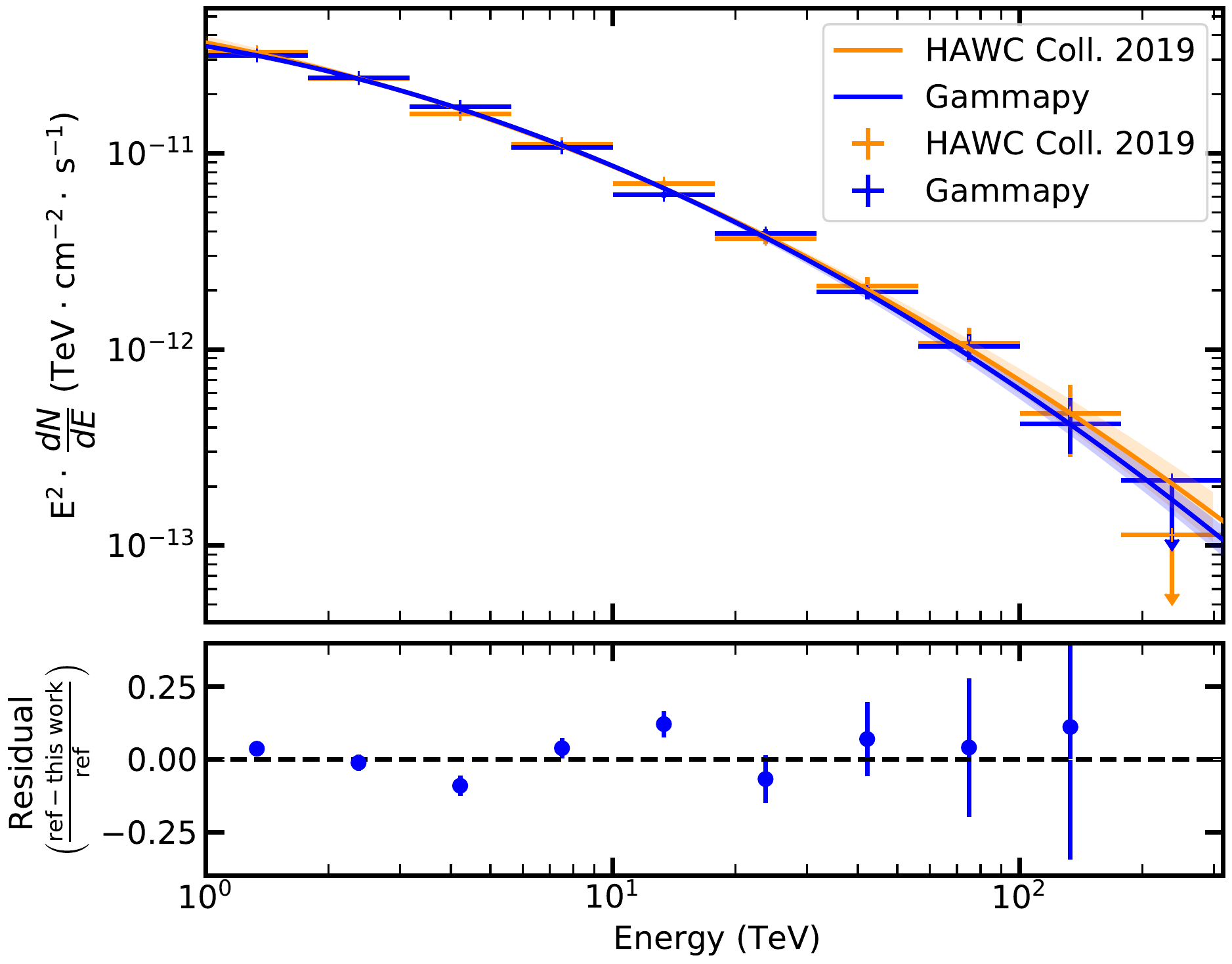}
        \caption{Best-fit Crab spectrum obtained with \textit{Gammapy} compared with~\cite{hawc-crab} for the GP energy estimator. The bottom panel shows the comparison between the flux points computed in this work and those reported in the reference.}
        \label{fig:crab_spectrum}
\end{figure}

As one of the brightest sources in the \gammaray sky, the Crab nebula is used as the standard candle for calibration and reference analysis. Due to its declination, it transits over the \hawc sky passing very close to the zenith. \hawc is able to detect (with a significance of roughly 5$\sigma$) the Crab nebula every day, that is, in the span of one transit.

We fit a point source and assumed a log-parabolic shape of the spectrum,
\begin{equation}
\label{eq:log parabola}
\frac{dN}{dE} = \phi_0 \left(E/E_0\right)^{-\alpha-\beta\ln\left(E/E_0\right)}~,
\end{equation}
where $E_0$ is the only fixed parameter with a value of 7~TeV.
We compared against~\cite{hawc-crab}. Figure~\ref{fig:crab_spectrum} shows the spectrum obtained with \gammapy and the exported data, compared against the reference analysis. The two results agree excellently.

\begin{table}
\begin{center}
\begin{tabular}{c c c c}
\hline  \hline 
~ & $\phi_0$ & $\alpha$ & $\beta$ \\
~ & \small{($10^{-13}$ TeV$^{-1}$ cm$^2$ s$^{-1}$)} & & \\
 \hline
\hline
From events& 2.39$\pm$0.04 & 2.79$\pm$0.02 & 0.113$\pm$0.007\\
Reference& 2.35$\pm$0.04 & 2.79$\pm$0.02 & 0.10$\pm$0.01\\
From map& 2.35$\pm$0.05 & 2.79$\pm$0.02 & 0.12$\pm$0.01\\
\hline
\end{tabular}
\end{center}
\caption{Maximum likelihood fit results for the Crab nebula. The fit result obtained using the DL3 products is given in the row labeled "From events". The fit result obtained using the standard HAWC map products is given in the "From map" row. The values in the "reference" column are taken from \cite{hawc-crab}.}
 \label{table:crab}
\end{table}
The resulting best-fit parameters are shown in Table~\ref{table:crab} as "From events", together with those from~\cite{hawc-crab}. Additionally, we repeated the exercise, but instead of using the exported data, we used the same standard HAWC counts and background map as were used in~\cite{hawc-crab}. The results of this fit are also shown in Table~\ref{table:crab} as "From map".

\begin{table*}
\begin{center}
\begin{tabular}{c c c c c c c}
 \hline   \hline

~ & R.A. & Dec & Extension (1$\sigma$) & $\phi_0$ & $\alpha$ & $\beta$ \\
~ & (\degr) & (\degr) & (\degr) & \small{($10^{-13}$ TeV$^{-1}$ cm$^2$ s$^{-1}$)} & & \\
 \hline
\hline
From events& 286.94$\pm$0.02 & 6.35$\pm$0.02 & 0.69$\pm$0.03 & 0.94$\pm$0.06 & 2.46$\pm$0.03 & 0.11$\pm$0.01\\
Reference & 286.91$\pm$0.10 & 6.32$\pm$0.09 & 0.67$\pm$0.03 & 0.95$\pm$0.05 & 2.46$\pm$0.03 & 0.11$\pm$0.02\\
From map & 286.96$\pm$0.03 & 6.36$\pm$0.03 & 0.68$\pm$0.03 & 0.94$\pm$0.06 & 2.45$\pm$0.04 & 0.12$\pm$0.02\\
\hline
\end{tabular}
\end{center}
\caption{Maximum likelihood fit results for eHWC J1907+063. The position in~\cite{hawc-he-catalog} is determined above 56~TeV, which leads to higher statistical errors due to the lower number of events. The fit result obtained using the DL3 products is given in the row labeled "From events". The fit result obtained using the standard HAWC map products is given in the "From map" row. The values in the "reference" column are taken from \cite{hawc-he-catalog}.}
\label{table:1908}
\end{table*}
\subsection{Extended source: eHWC J1907+063}
\label{subsec:j1908}

\begin{figure}[]
\centering
        \includegraphics[ width=0.46\textwidth]{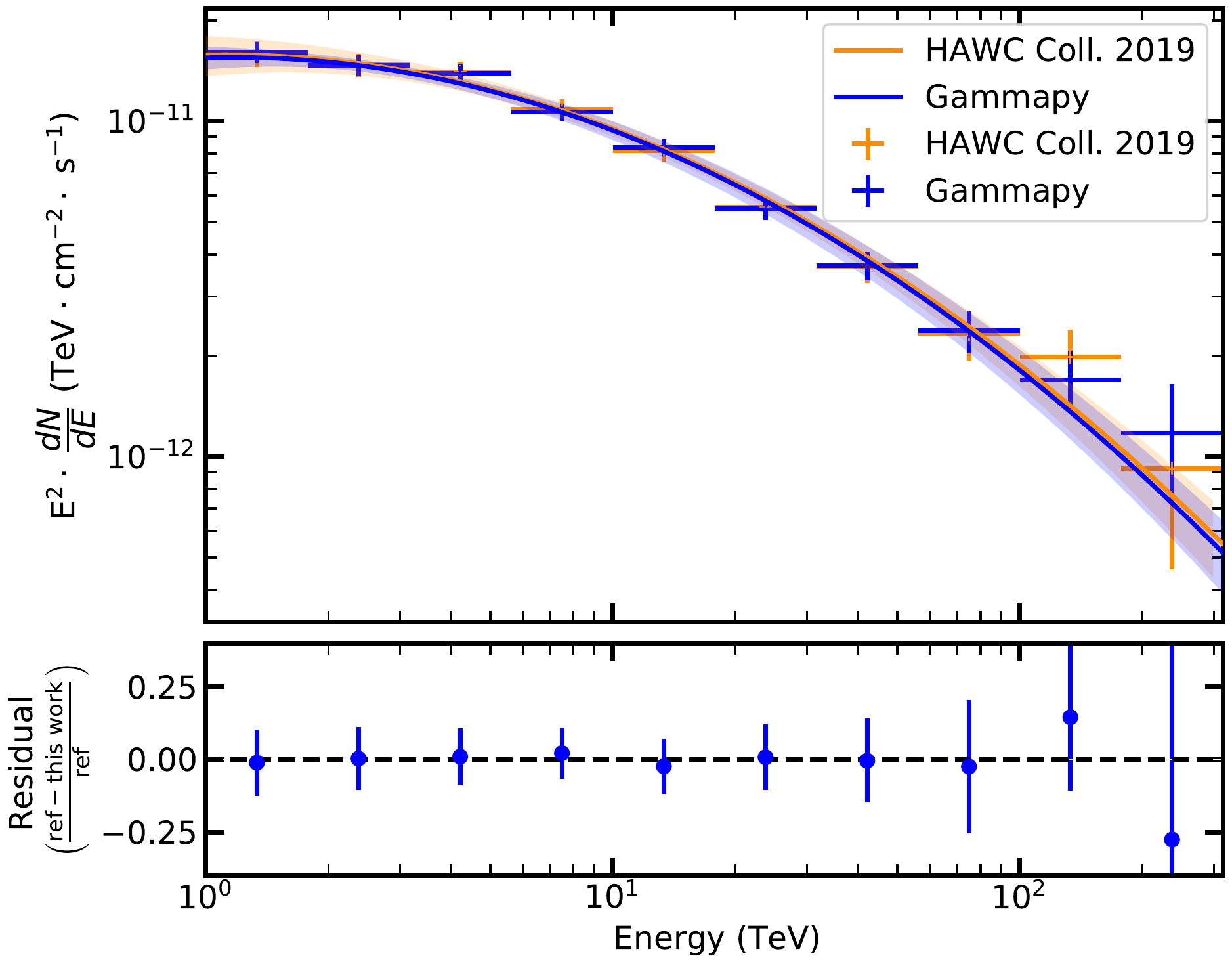}
        \caption{Best-fit spectrum of eHWC J1907+063 obtained with \textit{Gammapy} compared with~\cite{hawc-he-catalog}. The bottom panel shows the comparison between the flux points computed in this work and those reported in the reference. }
    \label{fig:j1908_spectrum}
\end{figure}

\begin{figure}
\centering
        \includegraphics[width=0.4\textwidth]{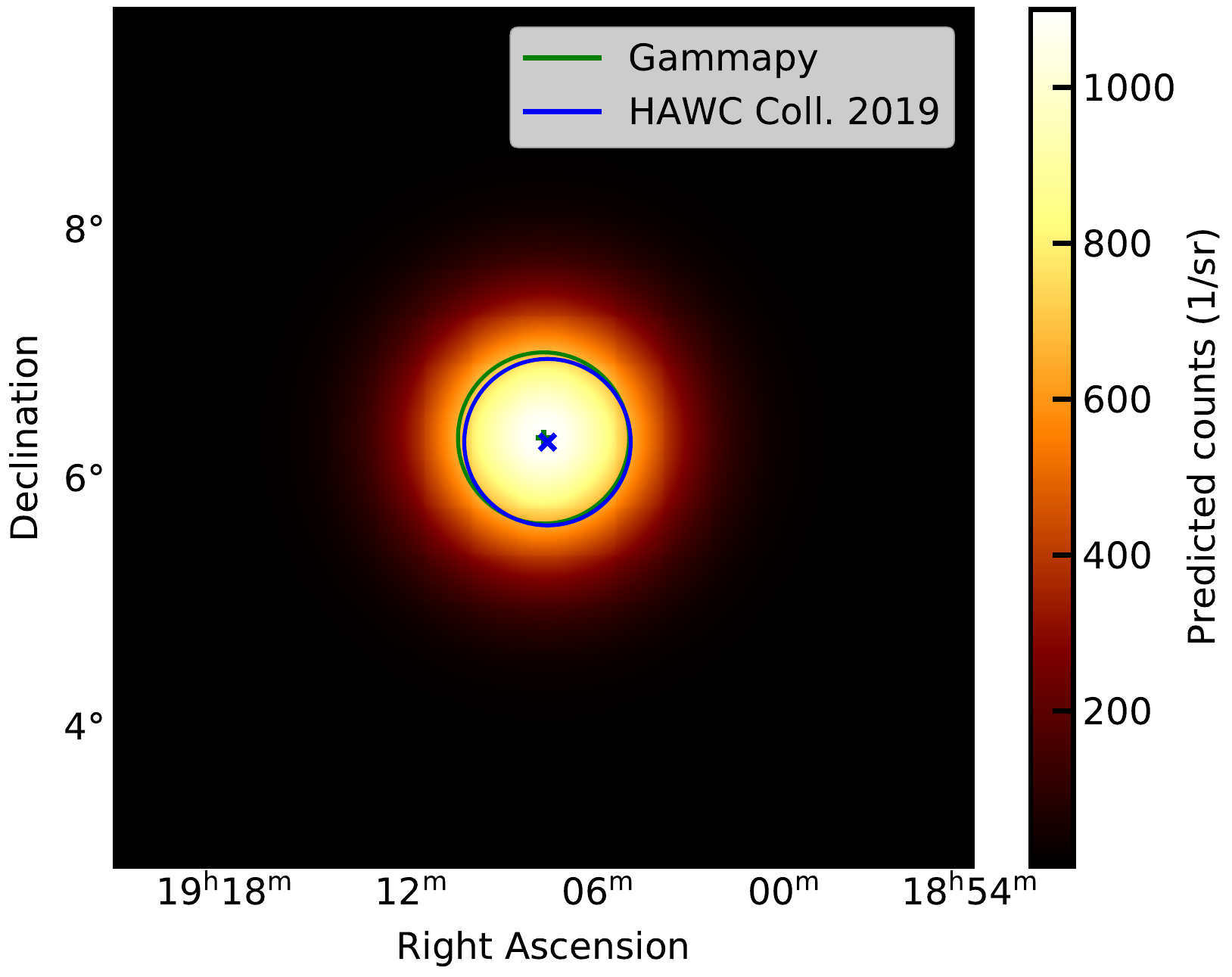}
        \caption{Spatial model of eHWC J1907+063 as obtained with \textit{Gammapy}. The green star and circle represent the best-fit position and the 68\% containment region, respectively. The blue cross and circle are the reference values from~\cite{hawc-he-catalog} for each quantity.}
                \label{fig:j1908_spatial}
\end{figure}

\cite{hawc-he-catalog} reported the detection by \hawc of several sources emitting above 56 and 100 TeV. One of those detected above 100 TeV is eHWC J1907+063. It is found in the vicinity of MGRO J1908+063. The 1$\sigma$ extension of the emission is reported to be 0.67\degr \ over the entire energy
range with a Gaussian assumption. The best-fit spectrum is modeled as a log-parabola (see Equation~\ref{eq:log parabola}), with the pivot energy $E_0$ fixed at 10~TeV.
We fit a combined spatial and spectral model made with the same assumptions as described above. Both components were fitted at the same time, including the source extension and position. The best-fit parameters are presented in Table~\ref{table:1908}.  Figure~\ref{fig:j1908_spectrum} shows the spectrum of eHWC J1907+063
compared against the reference analysis. The agreement is clearly excellent.
Figure~\ref{fig:j1908_spatial} shows the resulting best-fit spatial model compared to the result in~\cite{hawc-he-catalog}. When the errors detailed in Table~\ref{table:1908} are taken into account, the agreement is very good.

Additionally, we repeated the exercise, but instead of using the DL3 products, we used the same standard HAWC counts and background map as were used in~\cite{hawc-he-catalog}. The results of this fit are also shown in Table~\ref{table:1908} as "From map".
\subsection{Time domain: Mrk 421}
\label{subsec:mrk421}
\begin{figure*}[]
    \centering
    \includegraphics[width=0.95\textwidth]{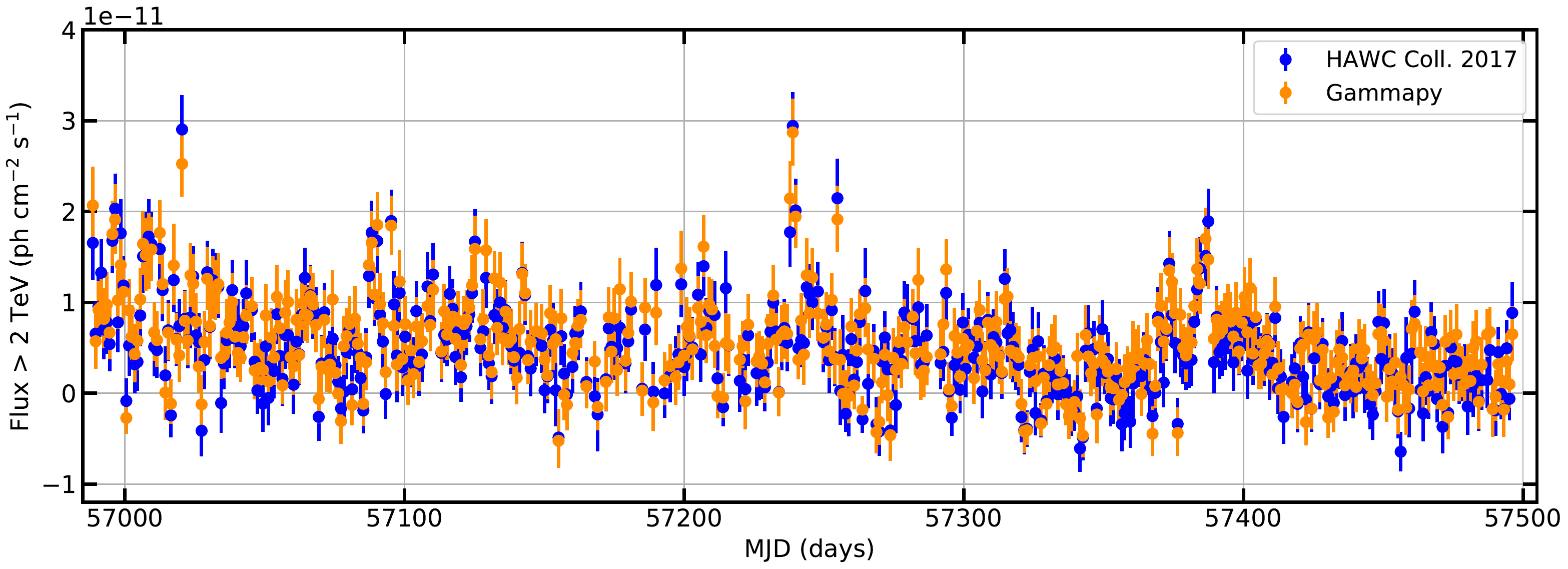}
        \caption{Light curve of Mrk 421 computed with \gammapy (orange points) compared to the reference in~\cite{hawc-lightcurve} (blue points).}
        \label{fig:mk_421_lightcurve}
\end{figure*}

Markarian (Mrk) 421 is a BL Lac object that has been extensively observed in the \gammaray band~\citep{hawcmrkrecent}. Its emission is known to be variable on timescales of hours or less. \cite{hawc-lightcurve} presented the \hawc light curve of Mrk 421, spanning over 17 months between November 2014 and April 2016. This work was carried out before the energy estimator techniques described in~\cite{hawc-crab} were implemented, which means that the energy of individual events could not be obtained. This leads to a different data selection and binning based only on \nhit, like the one described in~\cite{ref_hawc_crab_paper_2017}. Consequently, there is no such thing as an energy dispersion matrix for each of the \nhitbins. In order to deal with this, we introduced an assumed energy axis with a single bin for each \nhitbin dataset. This workaround allowed us to use the \gammapy framework in the same way as the previous two cases. 
The data selection and time binning were performed in a similar way as in~\cite{hawc-lightcurve}. Each event was associated with a sidereal day, starting at midnight  local  sidereal  time  at  the  \hawc  site. The current \hawc detector stability criteria for data selection were applied, noting that these are slightly stricter than those used by~\cite{hawc-lightcurve}. For each of the sidereal days, the fraction of a Mrk 421 transit that is included in the data was computed by integrating the curve shown in Figure~1 of~\cite{hawc-lightcurve}. Sidereal days for which this fraction is lower than 0.5 were removed from the selection. The result is a total of 463 transits, slightly fewer than the 471 included in~\cite{hawc-lightcurve} due to the stricter data selection cuts. For each of these transits, the background was estimated locally. This was done by masking the expected source location and computing the number of counts outside the mask in overlapping declination bands, which takes into account the varying instrument response with declination. Because Mrk 421 is seen by HAWC as an isolated point source, this approximation suffices. Finally, counts and background maps were bundled with the PSF and effective area, the latter corrected for the transit fraction.

A point source spatial model was used with the position fixed to (166.11\degr, 38.21\degr) in equatorial coordinates, as was done in~\cite{hawc-lightcurve} and~\cite{hawc-docu}. The spectrum of Mrk 421 was modeled by a power law with normalization $\phi_0$ at $E_0 = $1~TeV, photon index $\Gamma$, and an exponential cut-off $E_C$,
\begin{equation}
    \frac{dN}{dE} = \phi_0 \left( \frac{E}{\mathrm{E_0}}\right)^{-\Gamma} \exp{\left(-\frac{E}{E_C}\right)}.
\end{equation}
The best-fit spectrum was first obtained for the entire data range. The resulting parameters are presented in Table~\ref{table:mrk} together with those reported in~\cite{hawc-lightcurve}. In order to ensure a stable fit,  a minimum value for $E_C = 0.1$~TeV was imposed because of the high
correlation between $\Gamma$ and $E_C$. 
\begin{table}[h!]
\begin{center}
\begin{tabular}{ c c c c}
 \hline    \hline                    
~ & $\phi_0$ & $\Gamma$ & $E_C$ \\
~ & \small{($10^{-11}$ TeV$^{-1}$ cm$^2$ s$^{-1})$} & & \small (TeV) \\
 \hline
\hline
This work & 2.67$\pm$0.16 & 2.20$\pm$0.09 & 5.2$\pm$2.6\\
Reference & 2.82$\pm$0.19 & 2.21$\pm$0.14 & 5.4$\pm$1.1 \\
\hline
\end{tabular}
\end{center}
\caption{Maximum likelihood fit results for Mrk 421.}
 \label{table:mrk}
\end{table}

\begin{figure}[h!]
        \includegraphics[width=0.48\textwidth]{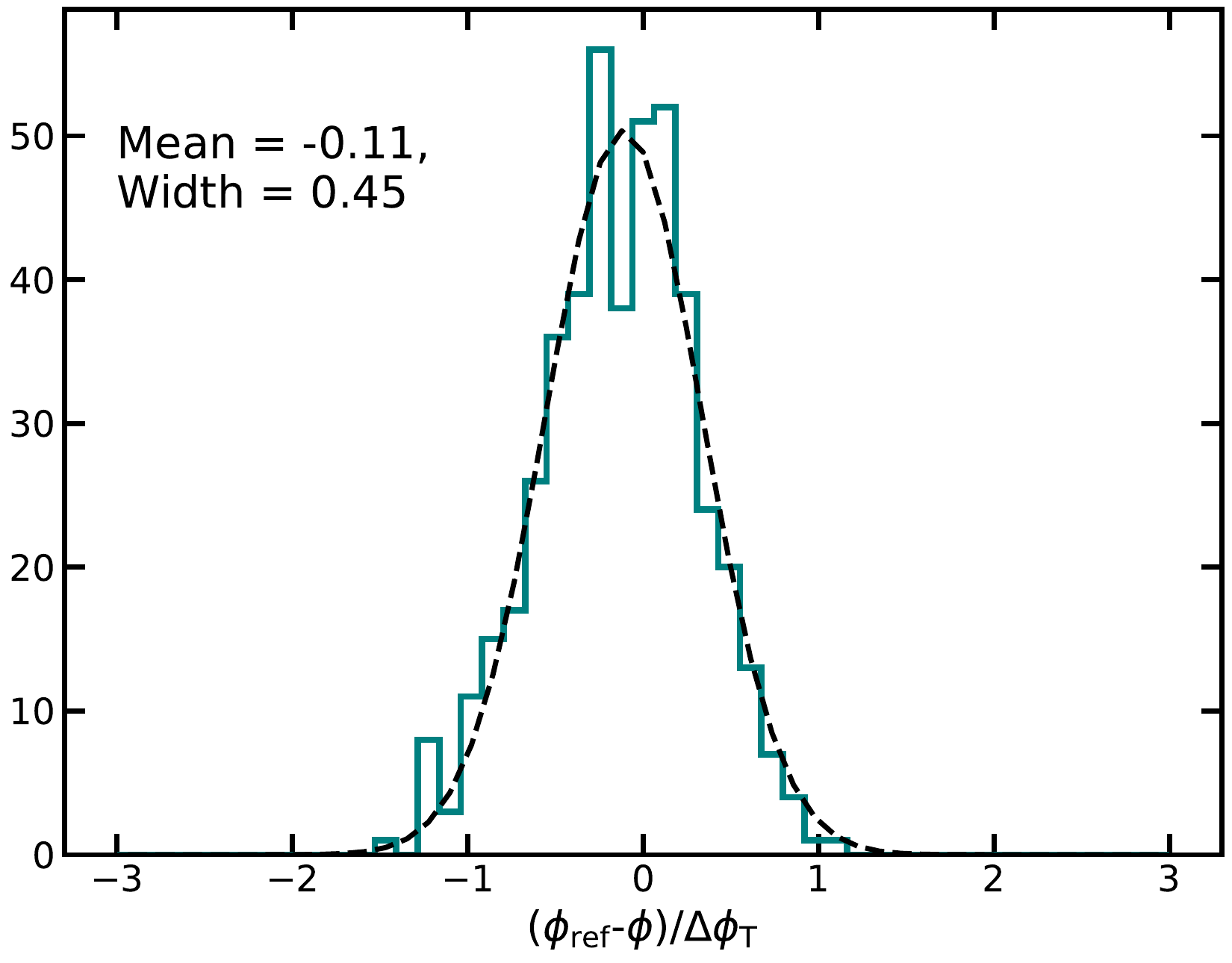}
        \caption{Distribution of the difference between the reference light-curve values ($\phi_{\mathrm{ref}}$) and those obtained with \gammapy ($\phi$) as a fraction of the combined error $\Delta \phi_{\mathrm{T}} =\sqrt{\Delta\phi_{\mathrm{ref}}^2 + \Delta\phi^2} $. A Gaussian fit is indicated with the dashed black line.}
        \label{fig:mrk421comparison}
\end{figure}

When the values of $E_C$ = 5~TeV and $\Gamma$ = 2.2 were fixed, the normalization was set free and was fit for each of the transits. The resulting spectra were integrated above 2~TeV to match the result of~\cite{hawc-lightcurve}. Figure~\ref{fig:mk_421_lightcurve} shows the light curve obtained with \gammapy together with the light curve from the reference analysis \citep{hawc-lightcurve}. The agreement is good, given the differences in data selection. The overall trend is clearly reproduced and the majority of points are compatible within errors. Figure~\ref{fig:mrk421comparison} shows the distribution of the differences between the reference light-curve points and those obtained with \gammapy as a fraction of the statistical error. The large majority of values clearly lies within the 1$\sigma$ region;the total is contained in the 2$\sigma$ region. 

\section{Proof of concept: Joint fit}
\label{sec:joint-fit}
\begin{figure}
        \includegraphics[width=0.48\textwidth]{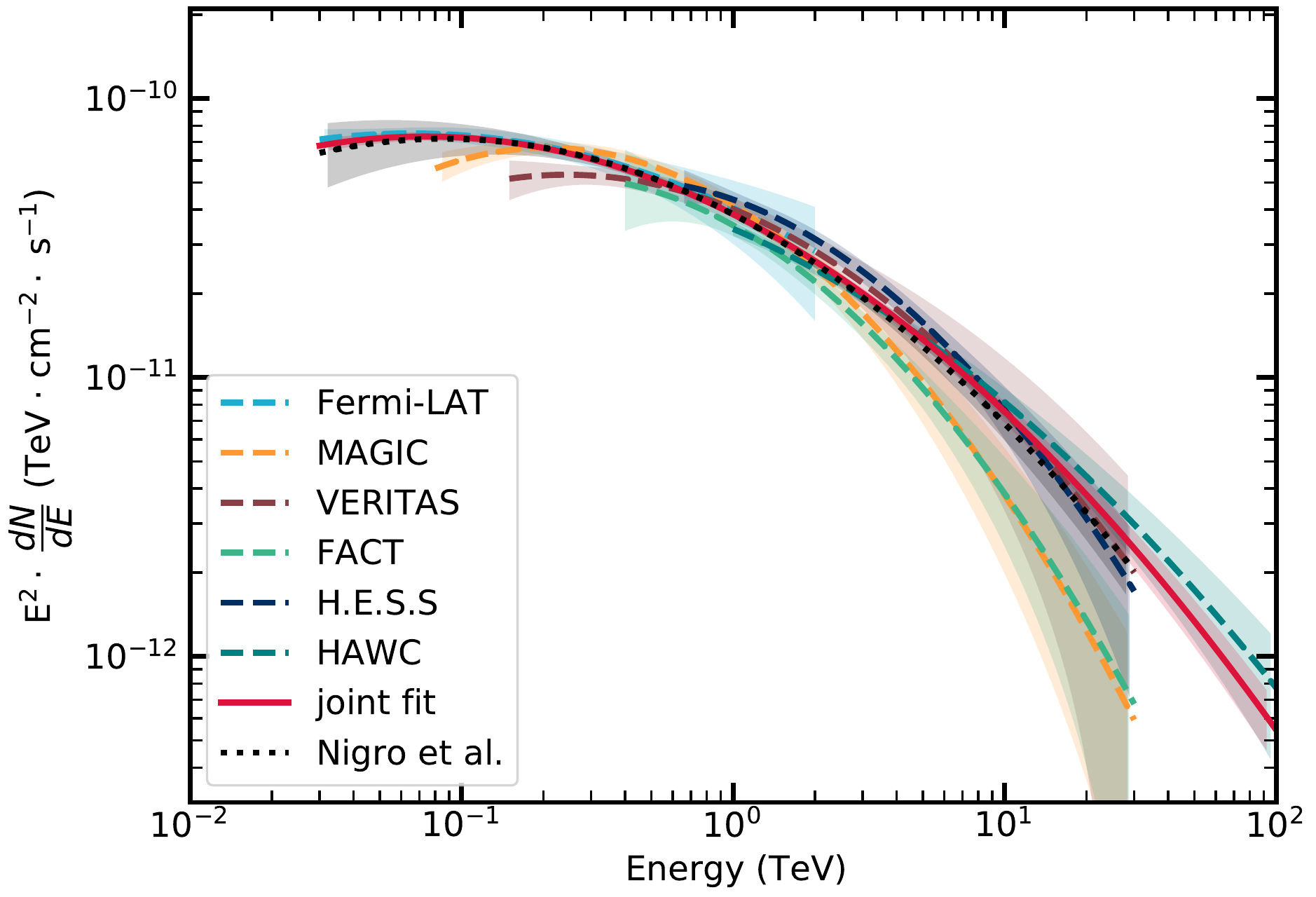}
        \caption{Crab nebula spectral energy distribution for individual instrument fits and from the joint fit. Single-instrument results are represented with dashed lines, and the fit of all the datasets together, labeled as joint, is represented as a thick solid red line. The joint fit result from~\cite{cosimo} is represented with a dotted black line.}
                \label{fig:jointfit}
\end{figure}

\cite{cosimo} presented the first fully reproducible measurement of the Crab nebula spectrum using public data from many different instruments. The analysis was carried out in \gammapy, and  emphasizes the power of a shared and open analysis tool. Similarly to~\cite{cosimo}, the goal of this study is not to reach any scientific conclusion regarding the Crab nebula. For this reason, we selected a small range of \hawc data, spanning only one month in time, and included it in the joint fit from~\cite{cosimo}. This was easily done due to the fully reproducible nature of that work. A log-parabola (see Equation~\ref{eq:log parabola}) spectral shape with fixed $E_0$ = 1~TeV was assumed for all the instruments. Performing the individual instrument data reductions and joint fits was straightforward after the data and \irfs were stored according to the \gadf. Figure~\ref{fig:jointfit} shows the result of this joint fit.

The spectrum of the Crab nebula might not be best described by the same spectral shape in all the different energy ranges, which could lead to differences in the best-fit parameters from the different experiments. However, this choice was made for simplicity, as the goal was not to reach any scientific conclusion regarding the Crab nebula, but rather show a proof of concept for this multi-instrument analysis. The joint fit shown in Figure~\ref{fig:jointfit} is indeed not noteworthy for the resulting spectral shape, but for the fact that a multi-instrument fit was performed using data from six different \gammaray instruments, including one satellite (\fermi), four \iacts, and one particle detector array (\hawc) natively within the same tool. 

The \hawc event lists and \irfs used in this section have been made publicly available on the \hawc Observatory website\footnote{\url{https://data.hawc-observatory.org/}}. This data release is the first to include \hawc data at the event list level. The data being public makes the result shown in Figure~\ref{fig:jointfit} fully reproducible, as an extension of~\cite{cosimo}.

\section{Conclusions and outlook}
\label{sec:conclusion}
We have presented the first full production of \hawc data and IRFs that follows the community-shared specifications of the GADF. Data in this format allow reusing existing high-level analysis tools such as \gammapy for the analysis of \hawc data. We validated this approach by reproducing several published \hawc results and found excellent agreement. We additionally reproduced the analysis using the maps that are typically produced by the \hawc Collaboration directly into \gammapy, which also yielded a very good agreement.

This cross-check does not only validate the analysis tools, it also provides a valuable cross-check of the corresponding \hawc results. The published results have been reproduced with high precision with a different analysis tool, which is a powerful, non-trivial check.

The lifetime of observatories is finite, and one of the concerns at the end of the operation is to ensure that the archival data are available and easy to use both for future studies and to reproduce previous results. In this regard, having data in a format that is shared across the community and that can be analyzed with a general-use tool is a key advantage.
The evolution of the \gadf will be driven by the requirements imposed by current and future observatories, which will require data and IRFs to be described in increasingly realistic and complex ways. This will directly benefit the current generation of instruments, which will be able to ensure that their legacy data are properly used and interpreted.

The joint Crab nebula fit presented in Section~\ref{sec:joint-fit} highlights the potential of this approach to perform multi-instrument analyses, spanning energy ranges much wider than those of a single instrument. This in turn can lead to synergies, bringing the \iact and particle detector array communities together. Future and current detectors, such as \swgo and \lhaaso, will  operate at the same time as CTA, and thus would benefit most from the ability to share data and analysis tools.
A shared analysis tool translates into a much larger developer and user base than any of the other collaboration-specific tools individually. This increases the complexity of features that can be implemented and maintained, benefiting all the instruments involved.

The work presented here is a proof of concept of what a particle detector array data analysis chain would look like when the shared format and tools are used. The very few limitations encountered arise because the initial development was led by the \iact community. However, these are minimal, and furthermore, expected to be resolved by future improvements, for example, with the expansion of the \gadf standard for sky maps, which are tremendously useful given the high event rates recorded by particle detector arrays. This development should be made taking existing standards into account when possible, and would allow data products to be efficiently distributed in map form as well.

The \gadf and science tools are constantly evolving to meet the needs of the community. Future particle detector arrays, such as \swgo, will be able to and should partake in this effort by ensuring that the format remains compatible with this detector class, while taking advantage of all the benefits it has to offer.

\begin{acknowledgements}
This work made use of \texttt{numpy}~\citep{numpy}, \texttt{scipy}~\citep{scipy}, \texttt{matplotlib}~\citep{matplotlib} and \texttt{astropy}~\citep{astropy}. We acknowledge the support from: the US National Science Foundation (NSF); the US Department of Energy Office of High-Energy Physics; the Laboratory Directed Research and Development (LDRD) program of Los Alamos National Laboratory; Consejo Nacional de Ciencia y Tecnolog\'ia (CONACyT), M\'exico, grants 271051, 232656, 260378, 179588, 254964, 258865, 243290, 132197, A1-S-46288, A1-S-22784, c\'atedras 873, 1563, 341, 323, Red HAWC, M\'exico; DGAPA-UNAM grants IG101320, IN111716-3, IN111419, IA102019, IN110621, IN110521; VIEP-BUAP; PIFI 2012, 2013, PROFOCIE 2014, 2015; the University of Wisconsin Alumni Research Foundation; the Institute of Geophysics, Planetary Physics, and Signatures at Los Alamos National Laboratory; Polish Science Centre grant, DEC-2017/27/B/ST9/02272; Coordinaci\'on de la Investigaci\'on Cient\'ifica de la Universidad Michoacana; Royal Society - Newton Advanced Fellowship 180385; Generalitat Valenciana, grant CIDEGENT/2018/034; Chulalongkorn University’s CUniverse (CUAASC) grant; Coordinaci\'on General Acad\'emica e Innovaci\'on (CGAI-UdeG), PRODEP-SEP UDG-CA-499; Institute of Cosmic Ray Research (ICRR), University of Tokyo, 
We also acknowledge the significant contributions over many years of Stefan Westerhoff, Gaurang Yodh and Arnulfo Zepeda Dominguez, all deceased members of the HAWC collaboration. Thanks to Scott Delay, Luciano D\'iaz and Eduardo Murrieta for technical support.

\end{acknowledgements}

%
%
\bibliographystyle{aa}

\bibliography{hawc-gammapy}
\end{document}